\RequirePackage{plautopatch}
\documentclass[reprint,amsmath,amssymb,aps,floatfix]{ptephy_v1}%%%%%% to generate preprint number with ptep logo
\usepackage[utf8]{inputenc} % umlaut input 
\usepackage[T1]{fontenc}    % umlaut encoding
\usepackage[dvipsnames,svgnames,psnames]{xcolor}
\usepackage[colorlinks,citecolor=DarkGreen,linkcolor=FireBrick,linktocpage,unicode]{hyperref}
\usepackage{graphicx,url,qrcode,bm,tikz,dsfont,slashed}

\usetikzlibrary{calc}
\usepackage{cleveref}
\usepackage{here}
\usepackage{subfig}
\usepackage{lscape}
\usepackage{dcolumn}

\Crefname{section}{Sec.}{Secs.}
\Crefname{section}{Section}{Sections}
\Crefname{figure}{Fig.}{Figs.}
\Crefname{equation}{Eq.}{Eqs.}
\Crefname{appendix}{Appendix}{Appendices}
\usepackage[normalem]{ulem}
\usepackage[arrowdel]{physics}
\newcommand{\red}{\color{black}}
\newcommand{\im}[0]{i}

\renewcommand{\vec}[1]{\bm{#1}}

\newcommand{\Slash}[1]{\ooalign{\hfil/\hfil\crcr$#1$}}% for Feynman slashs

\newcommand{\SUN}[1]{\text{SU} ( #1 )}

\newcommand{\MeV}{\mathrm{\ MeV}}
\newcommand{\GeV}{\mathrm{\ GeV}}

\newcommand{\timeope}{{\mathbf T}}

\begin{document}
\title{$K^+ N$ elastic scatterings for estimation of in-medium quark condensate with strange quarks}
\author{Yutaro Iizawa}
%\email{iizawa.y.aa@m.titech.ac.jp}
\author{Daisuke Jido}
\author{Stephan Hübsch}
% \affiliation{Department of Physics, Tokyo Institute of Technology, Meguro, Tokyo 152-8551, Japan}
\affil{Department of Physics, Tokyo Institute of Technology, Meguro, Tokyo 152-8551, Japan}
\date{\today}
\begin{abstract}
  We revisit the low-energy $K^+N$ elastic scatterings in the context of the in-medium quark condensate with strange quarks. The chiral ward identity connects the in-medium quark condensate to the soft limit value of the pseudoscalar correlation function evaluated in nuclear matter. The in-medium correlation function of the psuedoscalar fields with strangeness describes in-medium kaon propagation and is obtained by kaon-nucleon scattering amplitudes in the low density approximation. We construct the kaon-nucleon scattering amplitudes in chiral perturbation theory up to the next-to-leading order and add some terms  of the next-to-next-to-leading order with the strange quark mass to improve expansion of the strange quark sector. We also consider the effect of a possible broad resonance state around $P_\mathrm{lab} = 600\MeV/c$ for $I=0$ reported in the previous study. The low energy constants are determined by existent $K^+N$ scattering data. We obtain good reproduction of the $K^+p$ scattering amplitude by chiral perturbation theory, while the description of the $KN$ amplitude with $I=0$ is not so satisfactory due to the lack of low energy data. Performing analytic continuation of the scattering amplitudes obtained by chiral perturbation theory to the soft limit, we estimate the in-medium strange quark condensate. 
\end{abstract}
\maketitle

\section{Introduction}
Chiral symmetry is one of the symmetries of quantum chromodynamics (QCD) of the fundamental theory for strong interactions and is  broken dynamically in the low-energy as a phase transition phenomenon.
In the vacuum phase transition, the quark condensate $\langle \bar qq \rangle$
%${\langle \bar uu + \bar dd \rangle}$ 
is one of the order parameters for the symmetry breaking.
In the case of the exact chiral symmetry, the value of the quark condensate is zero when chiral symmetry is manifest, and after the symmetry breaking it becomes finite.
This is called the dynamical breaking of chiral symmetry (DB$\chi$S).
In extreme environments, e.g., high temperature and/or high density, the broken chiral symmetry is expected to be fully restored.
% Therefore, the broken chiral symmetry is expected to be fully restored at high temperature and/or high density. 

In order to confirm how DB$\chi$S takes place phenomenologically, we investigate the partial restoration of the chiral symmetry in nuclear matter. There
the magnitude of the quark condensate
%${\langle \bar uu + \bar dd \rangle}$ 
is expected to decrease as chiral symmetry is restored in nuclear matter. Since the quark condensate is not directly observable, it is necessary to obtain the information on the quark condensate through the experimental values of hadrons in nuclei, such as hadron-nucleus scatterings and bound states of hadrons in nuclei.
Of particular interest is the in-medium property of the Nambu-Goldstone boson (NG boson) such as pion.
The NG bosons appear to be associated with DB$\chi$S.
Hence the properties of the NG bosons should be sensitive to the nature of DB$\chi$S. The partial restoration of DB$\chi$S has been studied especially for pion in the nucleus, which extracts the in-medium quark condensate ${\langle \bar uu + \bar dd \rangle}$ for two flavors. 
From the observations of deeply bound pionic atoms \cite{Suzuki:2002ae} and the low-energy pion-nucleus elastic scatterings~\cite{Friedman:2004jh}, the isovector scattering length of $\pi$-nucleus system $b^*_1$ was extracted. Comparing $b^*_1$ and that in the $\pi N$ system $b_1$ based on theoretical considerations \cite{Kolomeitsev:2002gc,Jido:2008bk}, it is suggested that chiral symmetry is restored about 30\% at normal nuclear density.
Theoretically, the restoration of chiral symmetry in nuclear matter was predicted by a model-independent low-density theorem \cite{Drukarev:1988kd,Drukarev:1991fs}. In this relation, the sign of the experimental $\sigma_{\pi N}$ term or the theoretical $c_1$ parameter of chiral perturbation theory determines whether the magnitude of the quark condensate increases or decreases in nuclear matter. Since the $\sigma_{\pi N}$ term extracted from the experimental data of low-energy $\pi N$ scatterings~\cite{Gasser:1990ce,Alarcon:2011zs,Chen:2012nx,Hoferichter:2015dsa,Yao:2016vbz,RuizdeElvira:2017stg} is found to have a positive sign, the quark condensate should decrease in nuclear matter.
Similar results are obtained by in-medium chiral perturbation theory \cite{Kaiser:2007nv,Goda:2013bka,Hubsch:2021nih}, which is developed by Refs.~\cite{Oller:2001sn,Meissner:2001gz}.
% Therefore, systematic studies for other systems are needed to confirm the consistency of the partial recovery of chiral symmetry in other systems. However, it is not known whether partial recovery of chiral symmetry occurs systematically in systems other than pion-nucleus systems. For this reason, we would like to consider NG bosons with strangeness.

From the systematic point of view, we study the quark condensate with strange components in nuclear matter in this paper.
As with ${\langle \bar uu + \bar dd \rangle}$, a theoretical calculation for the low-density relation of ${\langle \bar uu + \bar ss \rangle}$ is performed based on the correlation function approach developed in Refs.~\cite{Jido:2008bk,Goda:2013bka}. There the in-medium quark condensate ${\langle \bar uu + \bar ss \rangle}$ is written in terms of kaon-nucleon scattering amplitude at the soft limit in the linear density approximation. We use chiral perturbation theory to extrapolate the scattering amplitude to the soft limit. The low energy constants (LECs) in the amplitude are determined from experimental data as well as the $\sigma_{\pi N}$ term.
% In-medium quark condensate with strange is calculated under the linear density approximation using in-medium chiral perturbation theory, and is obtained as a quantity including the low-energy constants (LECs) in the $\SUN{3}_f$ chiral perturbation theory as parameters \cite{StephanDron,IizawaHubschJido}
% \begin{align}\label{eq:uu_plus_ss_cond}
%   \frac{\langle \bar uu +\bar ss \rangle^*}{\langle \bar uu + \bar ss \rangle_0} & = 1 + \frac{(4b_0 + 3b_D - b_F)}{F_K^2}\rho
% \end{align}
% where $b_0$, $b_D$ and $b_F$ are the LECs of the next-to-leading order of $\SUN{3}$ chiral perturbation theory.
The current approach is a complementary method to the evaluation of the quark condensate by using the Feymann-Hellmann theorem where the quark condensate is obtained by taking the derivative of the nucleon energy density with respect to the quark mass~\cite{Oller:2001sn,Meissner:2001gz,Kaiser:2007nv,Kaiser:2008qu}.

We make good use of the $K^+ N$ scattering in order to determine the LECs. For determining the LECs, $K^+ N$ scatterings are preferable over $K^- N$ scatterings since in $\bar K N$ system the $\Lambda (1405)$ resonance appears below the threshold with a narrow decay width, while such a resonance does not exist in the $K^+$N system.
The $K^+ N$ scattering at low-energy has been studied for a long time \cite{Martin:1975gs,Martin:1980qe,Nakajima:1982tk,Hyslop:1992cs,Gibbs:2006ab}.
%The recent study is done by Refs.~\cite{Aoki:2018wug,Aoki:2017hel}.
Recently, Ref.~\cite{Aoki:2017hel} carried out the construction of the $K^+ N$ scattering amplitude using chiral perturbation theory up to the next-to-leading order, in which some terms were missing.
Reference~\cite{Aoki:2018wug} constructed the $K^+ N$ scattering amplitude using chiral unitary approach and discussed the presence of a broad resonance state with $I=0,\ S = +1$ around $P_\mathrm{lab} = 600\MeV/c$.
% It should be noted that the LECs in the chiral unitary approach are different from the original values as a result of the unitarization. 
For the purpose of determining the LECs, the $K^+ N$ scatterings need to be described by chiral perturbation theory. 
In our calculation, we construct $K^+ N$ scattering amplitudes using the chiral perturbation theory up to the next-to-leading order and some terms from the next-to-next order which includes the strange quark mass~\cite{Hyodo:2003qa} and determine the LECs using scattering data.
With the determined LECs we estimate the low-density behavior of the quark condensate with strange quarks in nuclear matter.

The structure of this paper is as follows. In Sec. \ref{sec:condensate}, we derive a relation to the in-medium quark condensate with strange quarks and the $KN$ scattering amplitude based on the correlation function approach \cite{Jido:2008bk,Goda:2013bka,Hubsch:2021nih} and the low-density theorem \cite{Drukarev:1988kd,Drukarev:1991fs}.
In Sec. \ref{sec:formulation}, we construct the $K^+ N$ scattering amplitudes using chiral perturbation theory.
In Sec. \ref{sec:results}, we determine the LECs so as to reproduce the existing $K^+ N$ scattering data.
% The total cross section and differential cross section data are compared with our results.
Using the determined LECs, we discuss the behavior of in-medium quark condensate with strange quarks. The quark condensates in hyperon matter and $\SUN{3}$ flavor symmetric baryonic matter are also discussed.
Moreover, we evaluate the wave function renormalization of the in-medium kaon.
In Sec. \ref{sec:conclusion}, we summarize the results of this paper.

\section{In-medium quark condensate with the strange quarks}\label{sec:condensate}
As mentioned in introduction, the purpose of this paper is to estimate the extended quark condensate to flavor $\SUN{3}$.
In this section, we describe the quark condensate in the nuclear medium based on the correlation function approach developed in Refs.~\cite{Jido:2008bk,Goda:2013bka,Hubsch:2021nih} and the low-density theorem~\cite{Drukarev:1988kd,Drukarev:1991fs}.
In this paper, we assume isospin-spin symmetric nuclear matter.

\subsection{Correlation function approach}\label{sec:ward}
Following Refs.~\cite{Jido:2008bk,Goda:2013bka,Hubsch:2021nih}, we calculate the divergence of the time-ordered product of the axial-vector current $A_\mu$ and the pseudoscalar field $P$ given as
\begin{align}\label{eq:ChWI}
  \partial^\mu \timeope\qty[A^\dagger_\mu (x) P(0)]
  = \timeope\qty[\partial^\mu A^\dagger_\mu (x) P(0)] + \delta(x_0) \comm{A^\dagger_0(x)}{P(0)}
\end{align}
where the axial-vector current $A_\mu(x)$ and the pseudoscalar field $P(x)$ are defined in terms of the up and strange quark fields as
\begin{align}
  A_\mu(x) & = \frac{1}{\sqrt{2}}\bar{s}(x)\gamma_\mu \gamma_5 u(x), \\
  P(x)     & = \sqrt{2} \im\bar{s}(x)\gamma_5 u(x),
\end{align}
respectively. The axial vector current is one of the Noether currents associated with the SU(3) chiral transformation and the pseusodcalar field appears in the partially-conserved axial current (PCAC) relation 
\begin{align}\label{eq:PCAC}
  \partial^\mu A_\mu(x) = \frac{m+m_s}{2} P(x),
\end{align}
with the explicit chiral symmetry breaking by the quark masses. Here $m$ and $m_s$ are the current quark masses of the light and strange quarks, respectively, with the isospin symmetry $m=m_u=m_d$. The pseudoscalar field is transformed under the axial transformation generated by $Q_5 \equiv \int d^3 x A_0^\dagger(x)$ as
\begin{align}
    [Q_5, P(x)] = -i S(x),
\end{align}
where the scalar field $S$ is given by
\begin{align} \label{eq:trans}
    S(x) = \bar u(x) u(x) + \bar s(x) s(x).
\end{align}

Evaluating \Cref{eq:ChWI} for the ground state of nuclear matter $\ket{\Omega}$ and introducing the in-medium correlation functions
\begin{align}
  \Pi_{5\mu}(x;\rho) & =\left\langle\timeope A_\mu ^\dagger(x) P(0)\right\rangle^*         \equiv \bra{\Omega}\timeope A_\mu ^\dagger(x) P(0)\ket{\Omega}, \\
  \Pi(x;\rho)        & = \left\langle\timeope P^\dagger(x) P(0)\right\rangle^* \equiv \bra{\Omega}\timeope P^\dagger(x) P(0)\ket{\Omega},
\end{align}
we obtain in the momentum space
%\begin{align}
%  \partial^\mu \Pi_{5\mu}(x,0;\rho) & = \frac{m+m_s}{2} \Pi(x,0;\rho) + \delta(x^0) \bra{\Omega}[A_0^\dagger(x), P(0)]\ket{\Omega}.
%\end{align}
%Performing the Fourier transformation to momentum space, we obtain
\begin{align}\label{eq:Fourier}
  % \delta^{ab}\langle \bar uu +\bar ss \rangle^* = -\im\frac{m+m_s}{2} \Pi^{ab}(q=0;\rho) \lim\limits_{q\to 0} \im \delta^{ab} \langle \bar uu +\bar ss \rangle^*
  -\im q^\mu \Pi_{5\mu}(q) & = \frac{m+m_s}{2} \Pi^{ab}(q) + \int \dd[3]{x} e^{-\im \vec{q}\cdot \vec{x}} \bra{\Omega}[A_0^\dagger(x), P(0)]\ket{\Omega}.
\end{align}
When we take the soft limit $q^\mu \to 0$ for \Cref{eq:Fourier}, the left-hand side vanishes since we do not have any zero modes off the chiral limit and the second term of the right-hand side yields the scalar field $S$ by using Eq.~\eqref{eq:trans}. 
%the in-medium quark condensate.
Finally, we have the in-medium condensate
\begin{align}\label{eq:cond_Pi}
  % \langle \bar uu +\bar ss \rangle^* & = -\im\frac{m+m_s}{4} \qty(\Pi^{44}(q=0;\rho)+\Pi^{55}(q=0;\rho))\nonumber                                                                       \\
  %                                    & = -\im\frac{m+m_s}{4} \left\langle P^4(0)P^4(0) + P^5(0)P^5(0) \right\rangle^*\nonumber                                                          \\
  %  & = -\im\frac{m+m_s}{2} \left\langle\qty(\frac{P^4(0) + \im P^5(0)}{\sqrt{2}}) \qty(\frac{P^4(0) - \im P^5(0)}{\sqrt{2}}) \right\rangle^*\nonumber \\
  \langle \bar uu +\bar ss \rangle^* & \equiv  \langle \Omega |\bar uu +\bar ss | \Omega \rangle  = -\im\frac{m+m_s}{2} \Pi(q = 0;\rho)
\end{align}
and the in-vacuum condensate
\begin{align}\label{eq:cond_Pi2}
  \langle \bar uu +\bar ss \rangle_0 & \equiv  \langle 0| \bar uu +\bar ss |0 \rangle = -\im\frac{m+m_s}{2} \Pi(q = 0;\rho = 0).
\end{align}

% Since the pseudoscalar filed $\frac{P^4(q) - \im P^5(q)}{\sqrt{2}}$ is related to the kaon filed $K^+(q)$ via $\frac{P^4(q) - \im P^5(q)}{\sqrt{2}} = G_K^{1/2} K^+(q)$ with the coupling constant $G_K^{1/2}$, we have
% \begin{align}\label{eq:propagator}
%   \Pi^{4+\im 5,4-\im 5}(q;\rho) & = G_K\bra{\Omega} K^-(q)K^+(q)\ket{\Omega}\nonumber             \\
%                                 & = \frac{\im G_K}{q^2 - M_K^2 - \Sigma_K(q;\rho) + \im \epsilon}
% \end{align}
% where $\Sigma_K(q)$ is the self-energy for in-medium kaon which stands for all of in-medium effects for the kaon and should be vanished at in-vacuum.
\subsection{Low-density theorem}
% Next we evaluate \Cref{eq:propagator} based on the low-density theorem \cite{Drukarev:1988kd,Drukarev:1991fs}.
% We perform an expansion for \Cref{eq:propagator} in terms of the density as
% \begin{align}\label{eq:LDE}
%   \Pi^{4+\im 5,4-\im 5}(q;\rho) & = \Pi^{4+\im 5,4-\im 5}(q;0) + \rho \eval{\pdv{\rho} \Pi^{4+\im 5,4-\im 5}(q;\rho)}_{\rho = 0} + O(\rho^{n>1})\nonumber                                                    \\
%                                 & = \frac{\im G_K}{q^2 - M_K^2 + \im \epsilon} + \rho \frac{\im G_K}{(q^2 - M_K^2 + \im \epsilon)^2} \eval{\pdv{\Sigma_K(q;\rho)}{\rho}}_{\rho = 0} + O(\rho^{n>1}\nonumber) \\
%                                 & = \Pi^{4+\im 5,4-\im 5}(q;0)\qty(1 + \frac{\rho}{q^2 - M_K^2 + \im \epsilon} \eval{\pdv{\Sigma_K(q;\rho)}{\rho}}_{\rho = 0} + O(\rho^{n>1})).
% \end{align}
% The self-energy $\Sigma_K(q;\rho)$ is also expanded in terms of the density as
% \begin{align}\label{eq:selfenergy}
%   \Sigma_K(q;\rho) & = \Sigma_K(q;0) + \rho\eval{\pdv{\Sigma_K(q;\rho)}{\rho}}_{\rho = 0} + O(\rho^{n>1})\nonumber \\
%                    & = \rho\eval{\pdv{\Sigma_K(q;\rho)}{\rho}}_{\rho = 0} + O(\rho^{n>1})
% \end{align}
% where we use the condition $\Sigma_K(q;0) = 0$.

In the low-density theorem \cite{Drukarev:1988kd,Drukarev:1991fs}, we can expand in-medium matrix element of an operator $\mathcal{O}$ as
\begin{align}\label{eq:LDT}
  \bra{\Omega}\mathcal{O}\ket{\Omega} = \bra{0}\mathcal{O}\ket{0} + \rho\bra{N}\mathcal{O}\ket{N} + O(\rho^{n>1}).
\end{align}
Applying this theorem to $\Pi(x;\rho)$, we obtain
\begin{align}\label{eq:LDT_K}
  \Pi(x;\rho) & = \bra{0} P^\dagger(x)P(0)\ket{0} + \rho \bra{N} P^\dagger(x)P(0)\ket{N} + O(\rho^{n>1}).
\end{align}
The matrix element $\bra{N} P^\dagger(x)P(0)\ket{N}$ is written by the isospin-averaged kaon-nucleon scattering amplitude $T_{K N}(q)$ using the reduction formula \cite{Weinberg:1966kf} as
\begin{align}\label{eq:reduction}
  \mathrm{F.T.}\bra{N} P^\dagger(x)P(0)\ket{N} = \frac{\im}{q^2-M_K^2} \frac{G^2_K}{q^2-M_K^2} \qty(-\frac{T_{KN}(q)}{2M_N})
\end{align}
where $G_K$ is the in-vacuum coupling defined as $\bra{0}P\ket{K^+} \equiv G_K$.
% Comparing \Cref{eq:LDE,eq:LDT_K} and using \Cref{eq:selfenergy,eq:reduction}, we have the $T\rho$ approximation for the self-energy for in-medium kaon as
% \begin{align}\label{eq:Trho}
%   \Sigma_K(q;\rho) = \rho \frac{T_{KN}(q)}{2M_N}.
% \end{align}
Finally, with \Cref{eq:cond_Pi,eq:cond_Pi2,eq:LDT_K,eq:reduction}, the quark condensate with strange components is given in terms of the isospin averaged kaon-nucleon scattering amplitude $T_{K N}$ in the soft limit  as
\begin{align}\label{eq:condensate}
  % \langle \bar uu +\bar ss \rangle^*                                             & = -\im\frac{m+m_s}{2} \Pi^{4+\im 5,4-\im 5}(q = 0;\rho)                                                                          \\
  %                                                                                & = -\im\frac{m+m_s}{2} \Pi^{4+\im 5,4-\im 5}(0;0) \qty(1 + \frac{\rho}{M_K^2} \frac{T_{KN}(q=0)}{2M_N}) \\
  \frac{\langle \bar uu +\bar ss \rangle^*}{\langle \bar uu + \bar ss \rangle_0} & = \qty(1 + \frac{\rho}{M_K^2} \frac{T_{KN}(q=0)}{2M_N}).
\end{align}
Here, in order to evaluate the condensate, it is necessary to take the soft limit for $T_{K N}$. For this purpose, $T_{K N}$ is constructed using chiral perturbation theory in the next section.

\section{Formulation for $K N$ amplitudes}\label{sec:formulation}
\subsection{The Chiral Lagrangian}\label{sec:Lagrangian}
In order to estimate the in-medium quark condensate with strange quarks, $\langle \bar uu + \bar ss \rangle^*$, based on \Cref{eq:condensate},
we construct the kaon-nucleon scattering amplitude $T_{K N}$  using chiral perturbation theory.
Chiral perturbation theory provides an analytic form of the scattering amplitude as a function of the energy and momentum. This is favorable for the analytic continuation of the scattering amplitude to the soft limit. The soft limit $q_\mu=0$ is not on the mass shell. Thus, the extrapolation to the soft limit has to be performed without taking the on-shell condition.
We determine the low-energy constants from the observed data of the $K^+ N$ scattering.
% We use chiral perturbation theory to describe the $K^+ N$ scattering amplitude. 
% The LECs are defined in chiral perturbation theory.

The leading order of the $\SUN{3}$ meson-baryon chiral Lagrangian reads
\begin{align}\label{eq:LO}
  \mathcal L_{MB}^{(1)} & = \Tr{\bar B (\im \slashed D - M_0) B)} - \frac{D}{2}\Tr{\bar B \gamma^\mu \gamma^5 \{u_\mu, B\}} - \frac{F}{2}\Tr{\bar B \gamma^\mu \gamma^5 [u_\mu, B]},
\end{align}
where $M_{0}$ is the baryon mass at the chiral limit, 
$D$ and $F$ are low energy constants to be determined by experiments, and
the baryon and meson fields, $B$ and $\Phi$, are written in the $\SUN{3}$ matrix form as
\begin{align}\label{eq:meson}
  \Phi = \begin{pmatrix}
           \frac{\pi^0}{\sqrt 2} + \frac{\eta}{\sqrt 6} & \pi^+                                         & K^+                    \\[8pt]
           \pi^-                                        & -\frac{\pi^0}{\sqrt 2} + \frac{\eta}{\sqrt 6} & K^0                    \\[8pt]
           K^-                                          & \bar K^0                                      & -\frac{2}{\sqrt 6}\eta
         \end{pmatrix},
\end{align}
\begin{align}\label{eq:baryon}
  B & = \begin{pmatrix}
          \frac{\Sigma^0}{\sqrt 2} + \frac{\Lambda}{\sqrt 6} & \Sigma^+                                            & p                         \\[8pt]
          \Sigma^-                                           & -\frac{\Sigma^0}{\sqrt 2} + \frac{\Lambda}{\sqrt 6} & n                         \\[8pt]
          \Xi^-                                              & \Xi^0                                               & -\frac{2}{\sqrt 6}\Lambda
        \end{pmatrix}.
\end{align}
Here we use the Coleman--Callan--Wess--Zumino (CCWZ) parametrization of the chiral field $U$ as
\begin{align}
  U=\exp(\im \sqrt{2}\Phi /f)
\end{align}
where $f$ is a normalization of the meson field $\Phi$ and corresponds to the meson decay constant at tree-level.
The covariant derivative for the baryon field is introduced as
\begin{align}
  D_\mu B = \partial_\mu B + \comm*{\Gamma_\mu}{B}
\end{align}
with the mesonic vector current given as
\begin{align}\label{eq:vector-current}
  \Gamma_\mu & = \frac{1}{2} (\xi^\dagger \partial_\mu \xi + \xi \partial_\mu \xi^\dagger)\nonumber
\end{align}
where $\xi^2=U$. The mesonic axial vector current is introduced as
\begin{align}
  u_\mu & = \im(\xi^\dagger \partial_\mu \xi - \xi \partial_\mu \xi^\dagger).
\end{align}
%The low-energy constants $D$ and $F$ are to be determined by fitting the semi-leptonic decays of the octet baryons at tree-level.

The next-to-leading order (NLO) of the chiral Lagrangian is given by
\begin{align}\label{eq:NLO}
  \mathcal L_{MB}^{(2)} & =  b_D \Tr{\bar B\{\chi_+, B\}} + b_F\Tr{\bar B[\chi_+,B]} + b_0 \Tr{\bar BB}\Tr{\chi_+}
  +d_1 {\rm Tr} \left(\bar{B} \{ u_{\mu}, [u^{\mu}, B] \} \right)\nonumber                                                    \\
                        & +d_2 {\rm Tr} \left(\bar{B} [u_{\mu}, [u^{\mu}, B]] \right)
  +d_3 {\rm Tr} \left(\bar{B} u_{\mu}){\rm Tr}(u^{\mu}B\right)
  +d_4 {\rm Tr} \left(\bar{B} B) {\rm Tr} (u^{\mu} u_{\mu} \right)\nonumber                                                   \\
                        & -\frac{g_1}{8M_N^2} {\rm Tr}  \left( \bar B \{ u_{\mu}, [ u_{\nu}, \{D^{\mu},D^{\nu}\}B] \} \right)
  -\frac{g_2}{8M_N^2} {\rm Tr}  \left( \bar B [u_{\mu}, [ u_{\nu}, \{D^{\mu},D^{\nu}\}B]] \right)\nonumber                    \\
                        & -\frac{g_3}{8M_N^2} {\rm Tr}  (\bar B  u_{\mu}) {\rm Tr}  ( u_{\nu}, \{D^{\mu},D^{\nu}\}B)
  -\frac{g_4}{8M_N^2} {\rm Tr}  (\bar B\{D^{\mu},D^{\nu}\}B) {\rm Tr} (u_{\mu} u_{\nu})\nonumber                              \\
                        & -\frac{h_1}{4} {\rm Tr}  \left( \bar B [\gamma^{\mu},\gamma^{\nu}]B u_{\mu} u_{\nu} \right)
  -\frac{h_2}{4} {\rm Tr}  \left(\bar B [\gamma^{\mu},\gamma^{\nu}] u_{\mu} [u_{\nu}, B] \right)\nonumber                     \\
                        & -\frac{h_3}{4} {\rm Tr}  \left(\bar B [\gamma^{\mu},\gamma^{\nu}] u_{\mu} \{u_{\nu}, B\} \right)
  -\frac{h_4}{4} {\rm Tr}  (\bar B [\gamma^{\mu},\gamma^{\nu}] u_{\mu}) {\rm Tr} (u_{\nu} B) + {\rm h.c.}
\end{align}
where $b_i$, $d_i$, $g_i$ and $h_i$ are the LECs of NLO.
The terms that include $b_i$ and $d_i$ appear in the typical flavor $\SUN{3}$ chiral Lagrangian such as in Ref.~\cite{Hyodo:2011ur,Aoki:2017hel}, while the terms that include $g_i$ and $h_i$ are introduced as the extension of the flavor $\SUN{2}$ chiral Lagrangian and used in Ref.~\cite{Aoki:2018wug}.
This Lagrangian is consistent with the most general form of the next-to-leading order shown in Refs.~\cite{Oller:2006yh,Geng:2013xn}.
% Among them $b_i$ appears in \Cref{eq:uu_plus_ss_cond} and determines the in-medium quark condensate $\langle \bar uu + \bar ss \rangle^*$. As we will see later, the combinations of these LECs in the isospin basis are determined by the $K^+ N$ elastic scatterings.

The scalar ($s=s^a\lambda^a$, $a=0,1,2\dots 8$) and pseudoscalar ($p=p^a\lambda^a$, $a=0,1,2\dots 8$) sources are contained in $\chi_\pm$ as
\begin{align}
  \chi_\pm & = \xi\chi^\dagger \xi \pm \xi^\dagger \chi \xi^\dagger
\end{align}
{\red through $\chi$ defined as}
%and the field $\chi$ contains the external scalar and pseudoscalar as
\begin{equation}
  \chi = 2 B_0 (s+\im p),
\end{equation}
where $B_0$ is a low-energy constant. The current quark masses are introduced through the external scalar field by setting 
\begin{align}
     s = \mathrm{diag}(m,m,m_s),    
\end{align}
with the isospin-averaged quark mass $m$ and the strange quark mass $m_s$. The low-energy constant $B_0$ is fixed with the current quark masses by the kaon mass with the relation $M_K^2 = B_0(m + m_s)$ in this work.

In order to improve extrapolation in the strange quark sector, we introduce some terms of the next-to-next-to-leading order (NNLO) of the chiral Lagrangian \cite{Oller:2006yh} which contain the strange quark mass $m_s$ in $\chi_-$ as
\begin{align}\label{eq:NNLO}
  \mathcal L^{(3)}_{MB} & = v_D \Tr (\bar B \acomm{\chi_-}{\gamma_5 B})
  + v_F \Tr (\bar B \comm{\chi_-}{\gamma_5 B})
  \nonumber                                                                                                                  \\
                        & + w_1 \Tr (\bar B \gamma_\mu B \comm{\chi_-}{u^\mu})
  + w_2 \Tr (\bar B \comm{\chi_-}{u^\mu} \gamma_\mu B )\nonumber                                                             \\
                        & +w_3 \qty[\Tr (\bar B u^\mu) \Tr(\chi_-\gamma_\mu B) - \Tr (\bar B \chi_-) \Tr(u^\mu\gamma_\mu B)]
\end{align}
where $v_i$ and $w_i$ are the LECs.
%%%
{\red There are other NNLO terms containing derivatives instead of the quark masses. As discussed in Ref.~\cite{Hyodo:2003qa}, mathematically the expansions in terms of the quark mass and the NG meson momentum are independent, although physically they are correlated through the Gell-Mann Oakes Renner relation. Here we would take the quark mass expansion more seriously.}
%%%

\subsection{Scattering amplitude}
To determine the LECs from the experimental data, we are allowed to take the on-shell condition on the external particles. In such a case,
the $T$-matrix for kaon and nucleon scattering is generally written as
\begin{align}\label{eq:Tmatrix1}
  T_{KN}(s,t) = \bar{u}(\vec{p}_4,s_4)\qty[A(s,t)+\frac{1}{2}(\Slash{p}_1+\Slash{p}_3)B(s,t)]u(\vec{p}_2,s_2),
\end{align}
where $p_1$ and $p_2$ denote the initial $K^+$ and nucleon momenta, respectively, while $p_3$ and $p_4$ stand for the final kaon and nucleon momenta, $u(\vec{p},s)$ is Dirac spinor with 3-momentum $\vec{p}$ and spin $s$, which is normalized by $\bar{u}(\vec{p},s)u(\vec{p},s^\prime) = 2M_N\delta_{ss^\prime}$ with nucleon mass $M_N$, and $A(s,t)$ and $B(s,t)$ are two Lorentz-invariant functions of the two independent Mandelstam variables $s = (p_1+p_2)^2$ and $t = (p_1-p_3)^2$.

The $K^+ N$ scattering amplitudes in the particle basis $T_{K^+ p\to K^+ p}$, $T_{K^+ n\to K^+ n}$ and $T_{K^+ n\to K^0 p}$ are constructed by those in the isospin basis $T^I\ (I=0,1)$ as
\begin{align}
  T_{K^+ p\to K^+ p} & = T^{I=1},                            \\
  T_{K^+ n\to K^+ n} & = \frac{1}{2}\qty(T^{I=1} + T^{I=0}), \\
  T_{K^+ n\to K^0 p} & = \frac{1}{2}\qty(T^{I=1} - T^{I=0}).
\end{align}

% Since the $K^+ N$ system is spin $0$ and spin $1/2$ system, 
Let us take the center-of-mass (c.m.) frame for partial wave decomposition. There we write the $T$-matrix in terms of non-spin-flip amplitude $f$ and spin-flip amplitude $g$ as
\begin{align}\label{eq:Tmatrix2}
  T(s,t) = \chi^\dagger(\lambda_4)\qty[f(W,\theta) - \im (\vec{\sigma}\cdot\vec{\hat{n}}) g(W,\theta)]\chi(\lambda_2)
\end{align}
where $W$ and $\theta$ are the total energy of the system and the scattering angle between $\vec{p}_1$ and $\vec{p}_3$ in the center-of-mass frame, respectively, $\vec{\hat{n}}$ is the normal vector of the scattering plane defined by
\begin{align}
  \vec{\hat{n}} = \frac{\vec{p}_3\times\vec{p}_1}{|\vec{p}_3\times\vec{p}_1|},
\end{align}
and $\chi(\lambda)$ is the Pauli spinor of a nucleon with helicity $\lambda$.

From \Cref{eq:Tmatrix1} and \Cref{eq:Tmatrix2}, we obtain the relation of the Lorentz-invariant amplitudes $A$, $B$ and the c.m. amplitudes $f$, $g$, as
\begin{align}
  f(W,\theta) & = (E_N + M_N)(A+\omega_K B) + k^2 B + \frac{(E_N + M_N + \omega_K)B-A}{E_N + M_N} k^2 \cos\theta, \\
  g(W,\theta) & = -\frac{(E_N + M_N + \omega_K)B-A}{E_N + M_N} k^2 \sin\theta,
\end{align}
where $E_N$, $\omega_K$ and $k$ stand for the nucleon energy, kaon energy and kaon momentum in the center-of-mass frame, respectively.
The amplitudes $f$ and $g$ are decomposed into the partial waves with Legendre polynomial $P_\ell(x)$ as
\begin{align}
  f(W,\theta) & = \sum_{\ell = 0}^{\infty} f_\ell (W) P_\ell (\cos\theta),                            \\
  g(W,\theta) & = \sum_{\ell = 0}^{\infty} g_\ell (W) \sin\theta\dv{P_\ell (\cos\theta)}{\cos\theta}.
\end{align}
We introduce the amplitude of the total angular momentum $j=\ell \pm \frac{1}{2}$, $T_{\ell\pm}$ as
\begin{align}
  f_{\ell}(W) & = (\ell+1) T_{\ell+}(W) + \ell T_{\ell-}(W), \\
  g_{\ell}(W) & = T_{\ell+} (W) - T_{\ell-}(W),
\end{align}
or equivalently
\begin{align}
  T_{\ell+}(W) & = \frac{1}{2\ell+1} (f_{\ell} (W) + \ell g_{\ell}(W)),     \\
  T_{\ell-}(W) & = \frac{1}{2\ell +1} (f_{\ell}(W) - (\ell+1) g_{\ell}(W)).
  \label{eq:amp_ell_pm}
\end{align}

By taking the average of the initial nucleon spins and the summation of the final nucleon spins, the differential cross section in the center-of-mass frame is calculated as
\begin{align}
  \dv{\sigma}{\Omega} = \frac{1}{64\pi^2 s} \qty(|f(W,\theta)|^2 + |g(W,\theta)|^2).
\end{align}
By integrating the differential cross section with respect to the solid angle $\Omega$, we obtain the total cross section as
\begin{align}
  \sigma = \frac{1}{32\pi s} \int \dd\cos\theta \qty(|f(W,\theta)|^2 + |g(W,\theta)|^2).
\end{align}

\subsection{$K^+ N$ scattering amplitude in chiral perturbation theory}\label{sec:ChPT}
In this section, we construct the tree-level amplitude of the $K^+ N$ elastic scattering using the chiral perturbation theory.
Here we consider the following four terms:
\begin{align}
  T^{K N} = T_\mathrm{WT} + T_\mathrm{Born} + T_\mathrm{NLO} + T_\mathrm{NNLO}.
\end{align}
The leading order contribution contains the amplitudes of the contact Weinberg-Tomozawa interaction $T_\mathrm{WT}$ and the $u$-channel Born terms of the hyperons $T_\mathrm{Born}$ with the $KYN$ Yukawa interactions given in \Cref{eq:LO}. The loop diagrams contribute from the next-to-next-to-leading order (NNLO).

% \subsubsection{Weinberg-Tomozawa term}
The invariant amplitudes for the Weinberg-Tomozawa diagram in the isospin basis are calculated from the leading order Lagrangian (\ref{eq:LO}) as
\begin{subequations}\label{eq:T_WT}
\begin{align}
  T^{I=0}_\mathrm{WT} &= 0,\\
  T^{I=1}_\mathrm{WT} &= \frac{1}{2F_K^2} \bar{u}(\vec{p}_4,s_4)(\Slash{p}_1+\Slash{p}_3)u(\vec{p}_2,s_2),
\end{align}    
\end{subequations}
and their corresponding invariant amplitudes read
\begin{subequations}\label{eq:WT}
\begin{align}
  A^{I=0}_\mathrm{WT} &= B^{I=0}_\mathrm{WT} = A^{I=1}_\mathrm{WT} =  0,\\
  B^{I=1}_\mathrm{WT} &= \frac{1}{F_K^2},
\end{align}    
\end{subequations}
with the kaon decay constant $F_K$.
% \subsubsection{Born term}
The invariant amplitudes for the $u$-channel Born terms in the isospin basis are evaluated as
\begin{subequations}\label{eq:T_Born}
\begin{align}
  T^{I=0}_\mathrm{Born} & = -\frac{3}{4}\frac{(D-F)^2}{F_K^2}
  \bar{u}(\vec{p}_4,s_4)\Slash{p}_1\gamma_5
  \frac{M_\Sigma + (\Slash{p}_2 - \Slash{p}_3) }{M_\Sigma^2 - (p_2 - p_3)^2 - i\epsilon}
  \Slash{p_3}\gamma_5 u(\vec{p}_2,s_2)
  \nonumber                                                                                              \\
                        & +\frac{1}{12} \frac{(3F+D)^2}{F_K^2}\bar{u}(\vec{p}_4 ,s_4)\Slash{p}_1\gamma_5
  \frac{M_\Lambda + (\Slash{p}_2 - \Slash{p}_3) }{M_\Lambda^2 - (p_2 - p_3)^2 - i\epsilon}
  \Slash{p_3}\gamma_5 u(\vec{p}_2,s_2),                                                                  \\
  T^{I=1}_\mathrm{Born} & = -\frac{1}{4}\frac{(D-F)^2}{F_K^2}
  \bar{u}(\vec{p}_4,s_4)\Slash{p}_1\gamma_5
  \frac{M_\Sigma + (\Slash{p}_2 - \Slash{p}_3) }{M_\Sigma^2 - (p_2 - p_3)^2 - i\epsilon}
  \Slash{p_3}\gamma_5 u(\vec{p}_2,s_2)
  \nonumber                                                                                              \\
                        & -\frac{1}{12} \frac{(3F+D)^2}{F_K^2}\bar{u}(\vec{p}_4 ,s_4)\Slash{p}_1\gamma_5
  \frac{M_\Lambda + (\Slash{p}_2 - \Slash{p}_3) }{M_\Lambda^2 - (p_2 - p_3)^2 - i\epsilon}
  \Slash{p_3}\gamma_5 u(\vec{p}_2,s_2)
\end{align}    
\end{subequations}
with the $\Sigma$ baryon mass $M_\Sigma$ and the $\Lambda$ baryon mass $M_\Lambda$.
The corresponding invariant amplitudes read
\begin{subequations}
\begin{align}
  A^{I=0}_\mathrm{Born} = & \frac34 \frac{(D-F)^2}{F_K^2}
  \frac{(M_N+M_\Sigma)(M_N^2-u)}{u - M_\Sigma^2}
  -\frac{1}{12} \frac{(3F+D)^2}{F_K^2}
  \frac{(M_N+M_\Lambda)(M_N^2 - u)}{u - M_\Lambda^2},      \\
  B^{I=0}_\mathrm{Born} = & -\frac34 \frac{(D-F)^2}{F_K^2}
  \frac{u+M_N^2 + 2 M_\Sigma M_N }{u - M_\Sigma^2}
  +\frac{1}{12} \frac{(3F+D)^2}{F_K^2}
  \frac{u+M_N^2 +2 M_\Lambda M_N}{u - M_\Lambda^2},        \\
  A^{I=1}_\mathrm{Born} = & \frac14 \frac{(D-F)^2}{F_K^2}
  \frac{(M_N+M_\Sigma)(M_N^2-u)}{u - M_\Sigma^2}
  +\frac{1}{12} \frac{(3F+D)^2}{F_K^2}
  \frac{(M_N+M_\Lambda)(M_N^2 - u)}{u - M_\Lambda^2},
  \quad                                                    \\
  B^{I=1}_\mathrm{Born} = & -\frac14 \frac{(D-F)^2}{F_K^2}
  \frac{u+M_N^2 + 2 M_\Sigma M_N }{u - M_\Sigma^2}
  -\frac{1}{12} \frac{(3F+D)^2}{F_K^2}
  \frac{u+M_N^2 +2 M_\Lambda M_N}{u - M_\Lambda^2},
\end{align}    
\end{subequations}
with the Mandelstam variable $u = (p_2 - p_3)^2$. We will use the isospin-averaged physical baryon masses for the calculation.

% \subsubsection{NLO}
The next-to-leading order of the $K^+N$ scattering amplitudes for $I=0,1$ is calculated from Lagrangian~\eqref{eq:NLO} as
\begin{align}\label{eq:T_NLO}
  T^I_\mathrm{NLO} = & \qty[\frac{4B_0}{F_K^2}(\hat{m}+m_s)b^I+\frac{2}{F_K^2}  (p_1\cdot p_3) d^I+\frac{(p_2\cdot p_1)(p_2\cdot p_3) + (p_4\cdot p_1)(p_4\cdot p_3)}{2M_N^2 F_K^2} g^I]\nonumber
  \\
                     & \times \bar{u}(\vec{p}_4,s_4)u(\vec{p}_2, s_2)
  - \frac{h^I}{2F_K^2}  p_1^{\mu} p_3^{\nu}\,  \bar{u}(\vec{p}_4,s_4) [\gamma_{\mu}, \gamma_{\nu}]
  u(\vec{p}_2, s_2),
\end{align}
where we have introduced the LECs for the NLO in the isospin basis, $b^I$, $d^I$, $g^I$ and $h^I$, which are written in terms of the LECs, $b_i$, $d_i$, $g_i$ and $h_i$ appearing in \Cref{eq:NLO} as
\begin{subequations} \label{eq:LECsNLO}
\begin{align}
  b^{I=0} & = b_{0} - b_{F},             & b^{I=1} & = b_{0} + b_{D}, \label{eq:bparas}\\
  d^{I=0} & = 2d_{1} + d_{3} - 2d_{4},   & d^{I=1} & = -2d_{2} - d_{3} - 2d_{4},   \\
  g^{I=0} & =  2g_{1} + g_{3} - 2 g_{4}, & g^{I=1} & =  -2g_{2} - g_{3} - 2 g_{4}, \\
  h^{I=0} & = h_{1}+h_{2}+h_{3}+h_{4},   & h^{I=1} & = h_{1}-h_{2}-h_{3}-h_{4}.
\end{align}    
\end{subequations}
The corresponding invariant amplitudes to \Cref{eq:T_NLO} read
\begin{subequations}
\begin{align}
  A_{\rm NLO}^I & = \frac{4B_0}{F_K^2}(m+m_s)b^I
  +\frac{2}{F_K^2}  (p_1\cdot p_3) d^I
  \nonumber                                      \\ &
  +\frac{(p_2\cdot p_1)(p_2\cdot p_3) + (p_4\cdot p_1)(p_4\cdot p_3)}{2M_{N}^{2} F_K^2} g^I
  + \frac{p_1\cdot(p_2+p_4)}{F_K^2} h^I          \\
  B_{\rm NLO}^I & = -\frac{2 M_{N}}{F_K^2} h^I.
\end{align}
\end{subequations}
% With the LECs in the isospin basis the in-medium quark condensate is expressed as
% \begin{align}\label{eq:uu_plus_ss_cond}
%   \frac{\langle \bar uu +\bar ss \rangle^*}{\langle \bar uu + \bar ss \rangle_0} & = 1 + \frac{(3b^{I=1} + b^{I=0})}{F_K^2}\rho.
% \end{align}
% From the $K^+N$ scattering amplitudes only two $b_i$ LECs out of three can be determined. Here we make use of the value of the $c_1$ parameter, $c_1 = 0.59$ GeV$^{-1}$ [**], which is fixed by the low-energy $\pi N$ scattering thanks to the flavor $\SUN{3}$ symmetry.

The $K^+ N$ scattering amplitudes obtained by using the NNLO chiral Lagrangian \eqref{eq:NNLO} for $I=0, 1$ are given by
\begin{subequations}\label{eq:T_NNLO}
\begin{align}
  T^{I=0}_\mathrm{NNLO}     = & \frac{3(D-F)(v_D - v_F)M_K^2}{F^2_K} \times \nonumber \\
   & \qquad \bar u(\vec{p}_4,s_4) \qty(\gamma_5 \frac{M_\Sigma + \Slash{p}_2-\Slash{p}_3}{M_\Sigma^2-(p_2-p_3)^2}\Slash{p}_3\gamma_5-\Slash{p}_1\gamma_5 \frac{M_\Sigma + \Slash{p}_2-\Slash{p}_3}{M_\Sigma^2-(p_2-p_3)^2}\gamma_5) u(\vec{p}_2,s_2)    \nonumber   \\
  -                           & \frac{(D+3F)(v_D + 3v_F)M_K^2}{3F^2_K} \times \nonumber \\
    & \qquad \bar u(\vec{p}_4,s_4) \qty(\gamma_5 \frac{M_\Lambda + \Slash{p}_2-\Slash{p}_3}{M_\Lambda^2-(p_2-p_3)^2}\Slash{p}_3\gamma_5-\Slash{p}_1\gamma_5 \frac{M_\Lambda + \Slash{p}_2-\Slash{p}_3}{M_\Lambda^2-(p_2-p_3)^2}\gamma_5) u(\vec{p}_2,s_2)\nonumber \\
  -                           & \frac{4 (w_{1} + w_{2} - w_{3})M_K^2}{F_K^2} \bar u(\vec{p}_4,s_4)(\Slash{p}_1 + \Slash{p}_3)u(\vec{p}_2,s_2),                                                                                                                                                                               \\
  T^{I=1}_\mathrm{NNLO}     = & \frac{(D-F)(v_D - v_F)M_K^2}{F^2_K} \times \nonumber \\
    & \qquad \bar u(\vec{p}_4,s_4) \qty(\gamma_5 \frac{M_\Sigma + \Slash{p}_2-\Slash{p}_3}{M_\Sigma^2-(p_2-p_3)^2}\Slash{p}_3\gamma_5-\Slash{p}_1\gamma_5 \frac{M_\Sigma + \Slash{p}_2-\Slash{p}_3}{M_\Sigma^2-(p_2-p_3)^2}\gamma_5) u(\vec{p}_2,s_2)    \nonumber    \\
  +                           & \frac{(D+3F)(v_D + 3v_F)M_K^2}{3F^2_K} \times \nonumber \\
    & \qquad \bar u(\vec{p}_4,s_4) \qty(\gamma_5 \frac{M_\Lambda + \Slash{p}_2-\Slash{p}_3}{M_\Lambda^2-(p_2-p_3)^2}\Slash{p}_3\gamma_5-\Slash{p}_1\gamma_5 \frac{M_\Lambda + \Slash{p}_2-\Slash{p}_3}{M_\Lambda^2-(p_2-p_3)^2}\gamma_5) u(\vec{p}_2,s_2)\nonumber \\
  -                           & \frac{4 (w_{1} - w_{2} + w_{3})M_K^2}{F_K^2} \bar u(\vec{p}_4,s_4)(\Slash{p}_1 + \Slash{p}_3)u(\vec{p}_2,s_2).
\end{align}    
\end{subequations}
These amplitudes are quark mass corrections of the Weinberg-Tomozawa interaction and $u$-channel Born terms.
The corresponding invariant amplitudes to \Cref{eq:T_NNLO} read
\begin{subequations}
\begin{align}
  A^{I=0}_\mathrm{NNLO} =   & \frac{6(D-F)v_-M_K^2}{F^2_K} \frac{(p_1\cdot p_4)+(p_2\cdot p_3)-M^2_K}{u-M_\Sigma^2}\nonumber    \\
  -                         & \frac{2(D+3F)v_+M_K^2}{3F^2_K} \frac{(p_1\cdot p_4) + (p_2\cdot p_3)-M^2_K}{u-M_\Lambda^2}        \\
  B^{I=0}_\mathrm{NNLO}  =  & -\frac{6(D-F)v_-M_K^2}{F^2_K} \frac{M_N + M_\Sigma}{u-M_\Sigma^2}\nonumber                        \\
  +                         & \frac{2(D+3F)v_+M_K^2}{3F^2_K} \frac{M_N + M_\Lambda}{u-M_\Lambda^2}-\frac{8M_K^2 w^{I=0}}{F_K^2} \\
  A^{I=1}_\mathrm{NNLO}   = & \frac{2(D-F)v_-M_K^2}{F^2_K} \frac{(p_1\cdot p_4) + (p_2\cdot p_3)-M^2_K}{u-M_\Sigma^2}\nonumber  \\
  +                         & \frac{2(D+3F)v_+M_K^2}{3F^2_K} \frac{(p_1\cdot p_4) + (p_2\cdot p_3)-M^2_K}{u-M_\Lambda^2}        \\
  B^{I=1}_\mathrm{NNLO}   = & -\frac{2(D-F)v_-M_K^2}{F^2_K} \frac{M_N + M_\Sigma}{u-M_\Sigma^2}\nonumber                        \\
  -                         & \frac{2(D+3F)v_+M_K^2}{3F^2_K}\frac{M_N + M_\Lambda}{u-M_\Lambda^2}-\frac{8M_K^2 w^{I=1}}{F_K^2},
\end{align}
\end{subequations}
where we have introduced the LECs for the NNLO as
\begin{subequations} \label{eq:LECsNNLO}
\begin{align}
  v_-     & = v_D - v_F,       & v_+     & = v_D + 3v_F,      \\
  w^{I=0} & = w_1 - w_2 + w_3, & w^{I=1} & = w_1 + w_2 - w_3.
\end{align}    
\end{subequations}
The low-energy constants $v_\pm$ in the next-to-next-to-leading order of Lagrangian are included in both isospin channels.
% Hence we will determine $v_\pm$ with $I=1$ LECs using the experimental data $I=1$ channel since the experimental data of $I=1$ channel is less ambiguity than that of $I=0$.

Applying the isospin-averaged kaon-nucleon amplitude to \Cref{eq:condensate}, we obtain the quark condensate in terms of the LECs defined in chiral perturbation theory as
\begin{align}\label{eq:uu_plus_ss_cond}
  \frac{\langle \bar uu +\bar ss \rangle^*}{\langle \bar uu + \bar ss \rangle_0} & = \qty(1 + \frac{\rho}{2M_N M_K^2} \frac{3T^{I=1}(q=0)+T^{I=0}(q=0)}{4})\nonumber \\
                                                                                 & = 1 + \frac{(3b^{I=1} + b^{I=0})}{F_K^2}\rho.
\end{align}
This extrapolation to the soft limit has been done without imposing the on-shell condition of the external particles. The expressions of \Cref{eq:T_WT,eq:T_Born,eq:T_NLO,eq:T_NNLO} have been obtained without taking the on-shell condition.
Using this equation, the quark condensate can be estimated directly from the LECs determined from experiments within the linear density.

\subsection{Coulomb correlation}
\label{Sec:Coulomb}
For the $K^+ p$ amplitude, we need to introduce the Coulomb correlation in order to compare it with the experimental data.
Here we follow the prescription done in Refs.~\cite{Aoki:2017hel,Aoki:2018wug} originally given in Ref.~\cite{Hashimoto:1984th}.
The Coulomb amplitude is calculated as
\begin{align}
  f_{C} = -\frac{\alpha}{2 k v \sin^{2}(\theta/2)} \exp\left[ - i \frac{\alpha}{v}
    \ln \left(\sin^2 \frac{\theta}{2} \right)\right]
\end{align}
with the scattering angle $\theta$, the fine structure constant $\alpha$
and the $K^+ N$ relative velocity $v$ defined by
\begin{align}
  v = \frac{ k (E_K + E_p)}{E_{K}E_{p}}.
\end{align}
We add the Coulomb amplitude to the strong interaction amplitudes calculated by the chiral perturbation theory. In addition, we multiply the Coulomb phase shift factor $e^{2i\Phi_\ell}$ with
\begin{align}
  \Phi_{\ell} = \sum_{n=1}^{\ell} \tan^{-1} \frac{\alpha}{n v},
\end{align}
for $\ell >0$ $(\Phi_0 = 0)$ to the strong interaction amplitudes. Finally, we have the amplitude with the Coulomb correlations as
\begin{align}
  f^{K^{+}p} & = \sum_{\ell=0}^{\infty} \left[ (\ell+1) T_{\ell+}^{I=1}
    + \ell T_{\ell-}^{I=1} \right] e^{2i\Phi_{\ell}} P_{\ell}(\cos\theta)
  - 8 \pi \sqrt s f_{C},                                                \\
  g^{K^{+}p} & = \sum_{\ell=1}^{\infty}
  \left[T_{\ell+}^{I=1} - T_{\ell-}^{I=1}\right]    e^{2i\Phi_{\ell}}
  \sin\theta \dv{P_\ell (\cos\theta)}{\cos\theta}.
\end{align}

\subsection{Inclussion of resonance state}
A recent work \cite{Aoki:2018wug} proposed the presence of a broad resonance state in the $KN$ scattering with $S=+1$ and $I=0$ around $\sqrt s = 1650\MeV$.
In Ref.~\cite{Aoki:2018wug}, the authors paid close attention to a sudden increase of the $I=0$ total cross section around $P_{\rm lab}=450 \MeV/c$ seen in the experimental data \cite{Carroll:1973ux} (Carroll~1973). They constructed the $K^+N$ scattering amplitudes using the chiral unitary approach and the model parameters were determined using observed cross sections of the $K^+N$ elastic scattering up to $P_\mathrm{lab}=800\MeV/c$.
They found two best solutions for the $K^+N$ amplitude with $I=0$; in Solution 1 the $P_{01}$ amplitude provides a dominant contribution, while in Solution 2 both $P_{01}$ and $P_{03}$ amplitudes contribute to the cross section. The former solution is more consistent with the Martin partial wave analysis \cite{Martin:1975gs}. Having performed analytic continuation of the obtained amplitudes into the complex energy plane, they found a resonance state in each solution. Solution 1 provides a resonance with $1617 \MeV$ mass and $305\MeV$ width in the $P_{01}$ partial wave, while Solution 2 finds the resonance with $1678 \MeV$ mass and $463\MeV$ width in the $P_{03}$ partial wave. The resonance parameters are summarized in \Cref{tab:resonance}. We will call the resonance in the former solution $P_{01}$ resonance and the latter one $P_{03}$ resonance in this paper.

The resonance energies correspond to $P_\mathrm{lab} \sim 600\MeV/c$ in the $K^+ N$ scattering. Since these resonances have a large width, the resonance may contribute to the $I=0$ scattering amplitude in a wide range of the energy around $P_\mathrm{lab} \sim 600\MeV/c$. In addition, most of low-energy data for the $I=0$ cross section are in these energies. As the pole terms associated with resonances cannot be expressed in the perturbative expansion of energy, we take account of the resonance contribution explicitly into our amplitudes.
% In our analysis, we will consider three cases for the $I=0$ amplitude: without resonances, with the $P_{01}$ resonance and with the $P_{03}$ resonance. 
% In our analysis, we consider four cases for the fitting: Carroll~1973 is used as the $I=0$ total cross section (FIT~1), Bowen~1970 is used (FIT~2), Bowen~1970 is used with the $P_{01}$ resonance (FIT~3) and Bowen~1970 is used with the $P_{03}$ resonance (FIT~4). 
The resonance state is introduced to the $I=0$ amplitude by adding the following amplitude to the appropriate partial wave amplitude $T_{\ell=1 \pm}$ defined in \Cref{eq:amp_ell_pm}:
% only when we use Bowen~1970 as the $I=0$ total cross section since the resonance state was found using Bowen~1970 as the $I=0$ total cross section.
% In the cases with the resonance we add the following resonance amplitude $T^\mathrm{Pole}$ to the appropriate partial wave of the chiral perturbation amplitude:
\begin{align}
  T^\mathrm{Pole} & = \frac{g^2 k^2}{\sqrt{s}-W+\im \Gamma/2},
  \label{eq:pole}
\end{align}
where $k$ is the c.m. momentum of the $K^+N$ scattering, $W$ and $\Gamma$ are the mass and width of the resonance state, respectively, and $g$ is the coupling strength of the resonance state to the $K^+N$ $I=0$ channel.
% with the $K^+N$ cm momentum $k$, the resonance mass $W$, the width $\Gamma$ and the coupling strength to the $I=0$ channel of the resonance $g$. 
The values of the coupling strengths are obtained as the residue of the scattering amplitudes at the resonance positions \cite{AokiPrivateComm}.

\begin{table}[t]
  \centering
  \caption{Property of the broad resonance states in the $KN$ scattering with $S=+1$ and $I=0$ around $\sqrt s = 1650\MeV$ reported by Ref.~\cite{Aoki:2018wug}. The coupling strengths are obtained from the residue of the scattering amplitudes at the resonance positions \cite{AokiPrivateComm}.}
  \begin{tabular}{l l c c c} \hline
    Solution   & Resonance ($J^{P}$)                     & mass [MeV] & width [MeV] & coupling strength [$10^{-3}\MeV^{-1}$] \\ \hline \hline
    Solution 1 & \enspace $P_{01}$ ($ \frac{1}{2}^{+}$)  & 1617       & 305         & $5.26 - 2.62\im$
    \\ \hline
    Solution 2 & \enspace $P_{03}$  ($ \frac{3}{2}^{+}$) & 1678       & 463         & $4.64 - 2.62\im$
    \\ \hline
    \label{tab:resonance}
  \end{tabular}
\end{table}

\section{Results}\label{sec:results}
In this section, we show the numerical results of our calculations. First of all, we determine the values of the LECs appearing in the scattering amplitudes from the existing $K^+N$ scattering data. We will see that the scattering amplitude for $I=1$ are constrained well by the $K^+ p$ elastic scattering data, while the scattering amplitude with $I=0$ is poorly determined due to the lack of data in particular for low energies and large ambiguity of the $I=0$ total cross section. Once the LECs are determined, we discuss the behavior of the quark condensate with strange quarks in the nuclear matter by using \Cref{eq:uu_plus_ss_cond}. We also discuss the in-medium quark condensates in hypothetical hyperonic matter in the view of the flavor symmetry. In addition, we show our calculation of the wave function renormalization of the in-medium kaon. We use the isospin-averaged hadron masses as summarized in \Cref{tab:FixedParam}.
 
\begin{table}[t]
  \centering
  \caption{Values of the physical constants we use.}
  \begin{tabular}{ccccccc}
    \hline
    $M_{N}$     & $M_{K}$     & $M_{\Lambda}$ & $M_{\Sigma}$ & $F_{K}$     & $D$  & $F$  \\
    \hline\hline
    $938.9$ MeV & $495.6$ MeV & $1115.7$ MeV  & $1193.2$ MeV & $110.0$ MeV & 0.80 & 0.46 \\
    \hline\label{tab:FixedParam}
  \end{tabular}
\end{table}

\subsection{Determining LECs}
We use the values of the low-energy constants in the leading order of Lagrangian, $D$ and $F$, given in Ref.~\cite{Luty:1993gi}, which are fixed by the hyperon semi-leptonic decays at tree level. The explicit values are shown in \Cref{tab:FixedParam}. 

The values of the LECs for the NLO and NNLO, $b^I$, $d^I$, $g^I$, $h^I$, $w^I$ for $I=0,1$ and $v_\pm$, given in Eqs.~\eqref{eq:LECsNLO} and \eqref{eq:LECsNNLO} are determined by carrying out the $\chi^{2}$ fitting of the $K^+ N$ amplitude obtained by chiral perturbation theory to the experimental data. The reduced $\chi^{2}$ function is defined as
\begin{align}\label{eq:chi-squared}
  \chi^{2}_\mathrm{d.o.f} = \frac{1}{{\cal N}_\mathrm{d.o.f}}\sum_{i}^{n} \left(\frac{y_{i} - f(x_{i})}{\sigma_{i}} \right)^{2}
\end{align}
where $y_{i}, f(x_{i}), \sigma_{i}$ and $n$ are the experimental data, the theoretical calculations with the parameters, the uncertainties of the data and the number of the data, respectively, and ${\cal N}_\mathrm{d.o.f.}$ stands for the number of degrees of freedom defined as ${\cal N}_\mathrm{d.o.f.} = n-m$ with the number of the LECs $m = 12$. In our calculation, we consider the partial waves up to the $D$-wave ($\ell=2$) in the theoretical amplitudes. We will check the convergence of the partial wave decomposition. We restrict the energy region up to $P_\mathrm{lab} = 800\MeV/c$, where inelastic contributions such as pion production start to be significant.

We determine all of the NLO and NNLO LECs simultaneously by using the experimental data of the $K^+ p$ differential cross section between $P_\mathrm{lab} = 145\MeV/c$ and $726\MeV/c$~\cite{Cameron:1974xx}, the $K^+ n\to K^0 p$ charge exchange differential cross sections between $P_\mathrm{lab} = 434\MeV/c$ and $780\MeV/c$~\cite{BGRT:1973bda,Damerell:1975kw}, the $I=1$ total cross section between $P_\mathrm{lab} = 145\MeV/c$ and $788\MeV/c$~\cite{Cameron:1974xx,Bugg:1968zz,Carroll:1973ux,Bowen:1970azd,Adams:1971ius,Bowen:1973ga}, and the $I=0$ total cross sections between $P_\mathrm{lab} = 413\MeV/c$ and $794\MeV/c$~\cite{Carroll:1973ux} and between $P_\mathrm{lab} = 366\MeV/c$ and $714\MeV/c$ \cite{Bowen:1970azd,Bowen:1973ga}. There are significant difference between the $I=0$ total cross sections given in Refs.~\cite{Bowen:1970azd,Carroll:1973ux}.
% As discussed in the previous section, we perform four different fittings; Carroll~1973 \cite{Carroll:1973ux} is used as the $I=0$ total cross section (FIT~1), Bowen~1970 \cite{Bowen:1970azd} is used (FIT~2), Bowen~1970 is used with the $P_{01}$ resonance (FIT~3) and Bowen~1970 is used with the $P_{03}$ resonance (FIT~4).

In this work, we consider four different fitting procedures for $I=0$: FIT~1 uses Carroll~1973~\cite{Carroll:1973ux} for the $I=0$ total cross section, while FIT~2 employs Bowen~1970 \cite{Bowen:1970azd}.
Both cases do not introduce the broad resonance into the $I=0$ amplitude.
FIT~3 considers the $P_{01}$ resonance by adding the resonance contribution (\ref{eq:pole}) to the $P_{01}$ scattering amplitude, while FIT~4 takes account of the $P_{03}$ resonance. In FIT~3 and FIT~4, we use Bowen~1970 for the $I=0$ total cross section, because the resonance properties were obtained by using Bowen~1970 in Ref.~\cite{Aoki:2018wug}.
In all four fittings, we do not use the differential cross sections of the $K^+ n$ elastic scattering due to their large experimental uncertainties.

The determined LECs for each case are summarized in \Cref{tab:LECs}.
The table shows that the values of LECs for $I=1$ in FIT~1, 2 and 4 are consistent with each other. 
We will see that a second best solution of FIT~3 is also consistent with these fits. 
This implies that the $K^+p$ experimental data constrain the $I=1$ $K N$ amplitude very well.

\begin{figure}[H]
  \centering
  \includegraphics[width=0.6\textwidth]{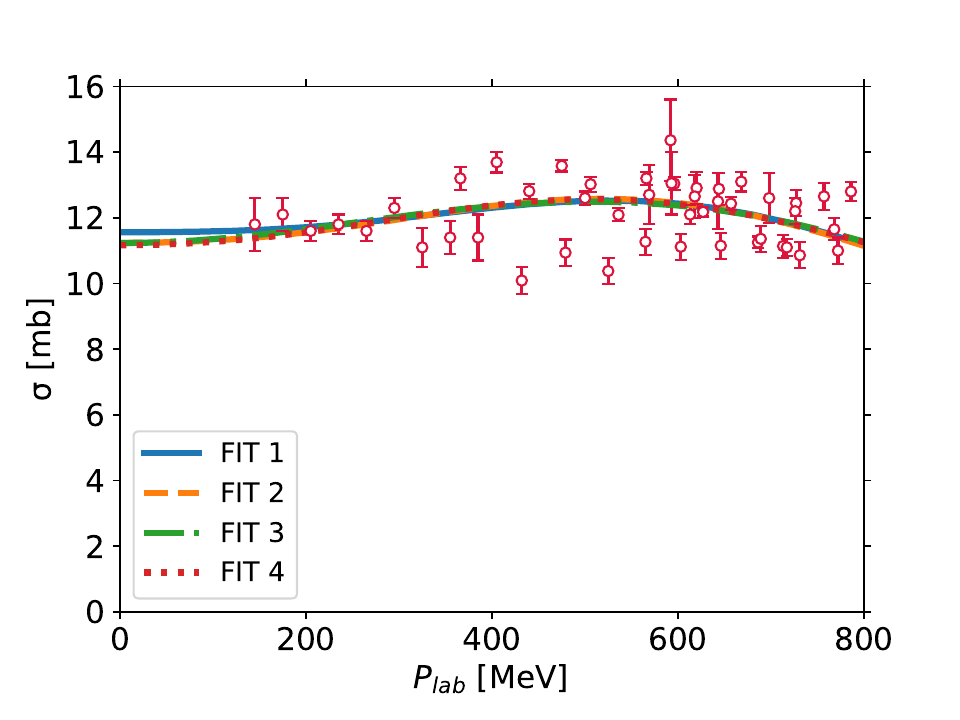}
  \caption{$I=1$ $K^+ N$ total cross sections calculated with the determined LECs given in Table.~\ref{tab:LECs} in comparison with the experimental data \cite{Cameron:1974xx,Bugg:1968zz,Carroll:1973ux,Bowen:1970azd,Adams:1971ius,Bowen:1973ga}.
    % The partial wave components are also described by the dashed line.
  }
  \label{fig:crosssection_I_1}
\end{figure}

\begin{figure}[H]
  \begin{tabular}{cc}
    \begin{minipage}[c]{0.5\hsize}
      \centering
      \includegraphics[width=0.65\textwidth]{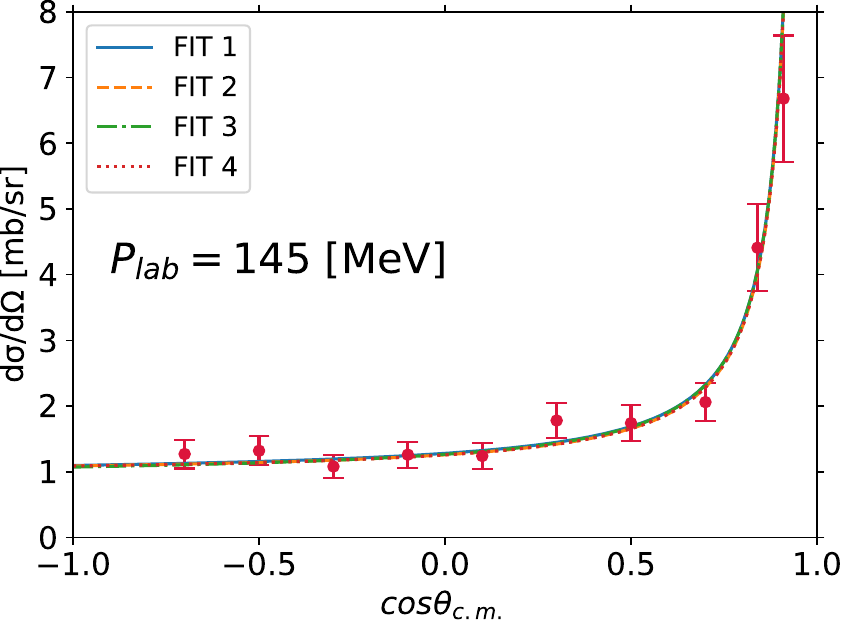}
    \end{minipage} &

    \begin{minipage}[c]{0.5\hsize}
      \centering
      \includegraphics[width=0.65\textwidth]{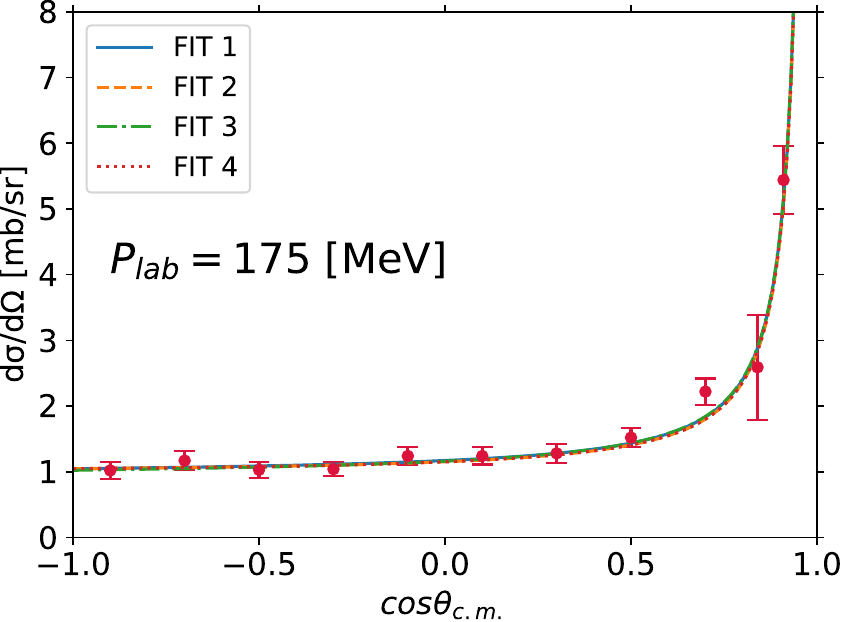}
    \end{minipage} \\

    \begin{minipage}[c]{0.5\hsize}
      \centering
      \includegraphics[width=0.65\textwidth]{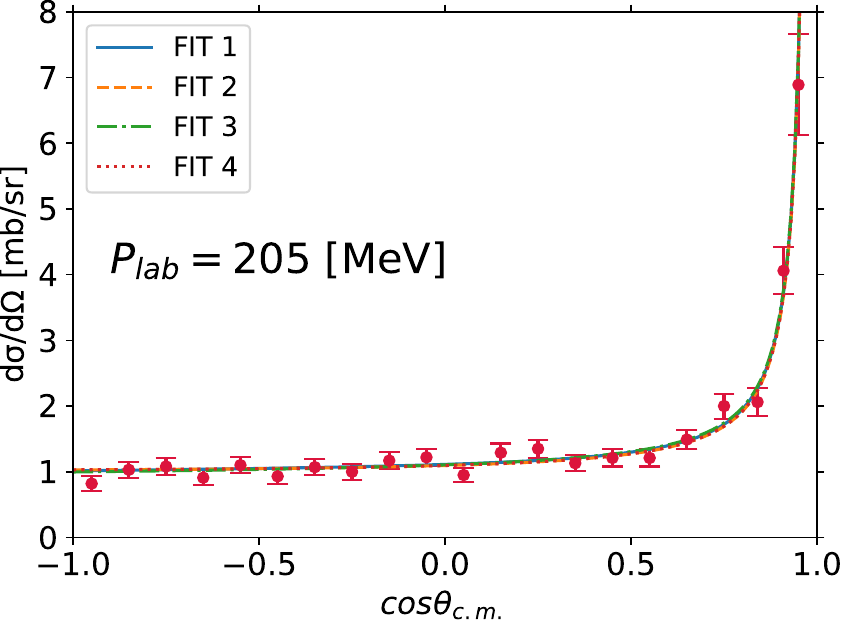}
    \end{minipage} &

    \begin{minipage}[c]{0.5\hsize}
      \centering
      \includegraphics[width=0.65\textwidth]{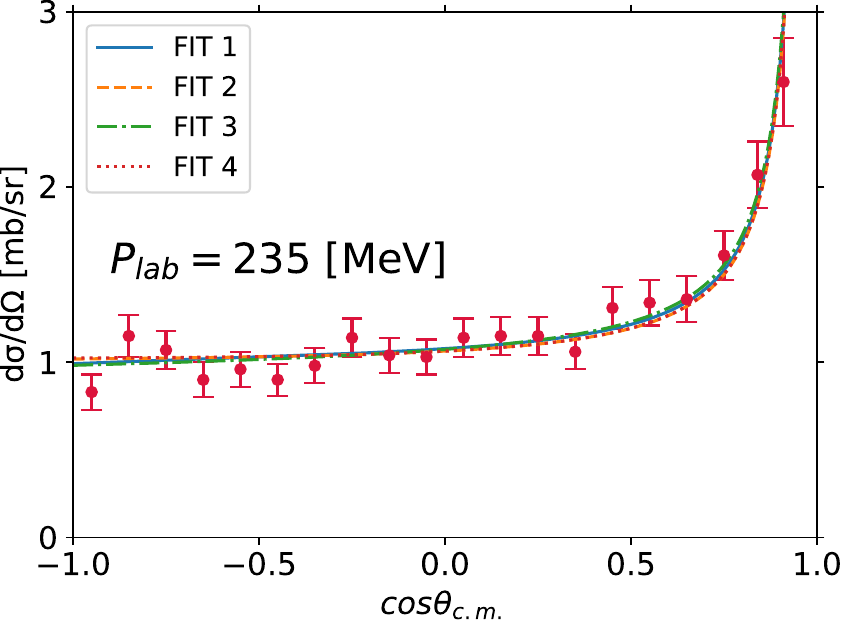}
    \end{minipage} \\
    \begin{minipage}[c]{0.5\hsize}
      \centering
      \includegraphics[width=0.65\textwidth]{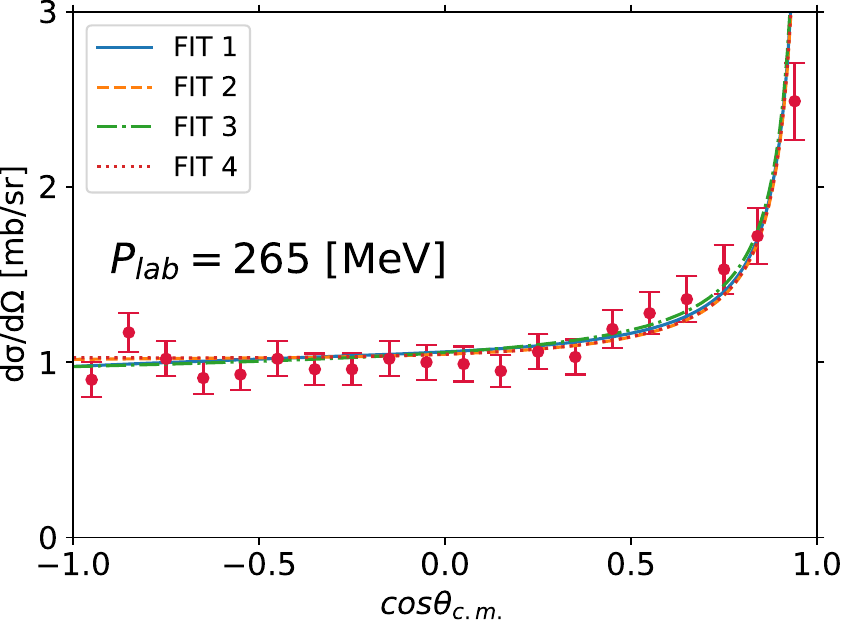}
    \end{minipage} &

    \begin{minipage}[c]{0.5\hsize}
      \centering
      \includegraphics[width=0.65\textwidth]{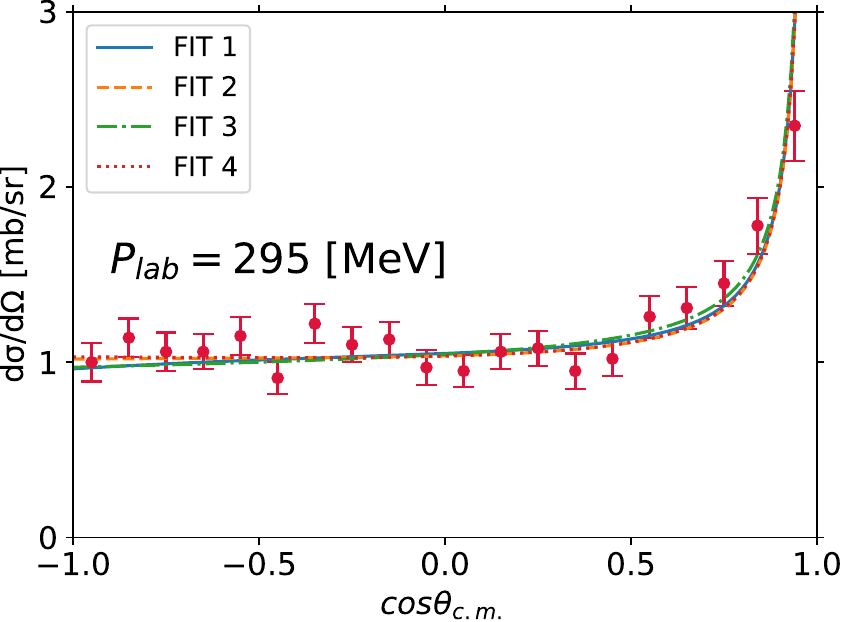}
    \end{minipage} \\

    \begin{minipage}[c]{0.5\hsize}
      \centering
      \includegraphics[width=0.65\textwidth]{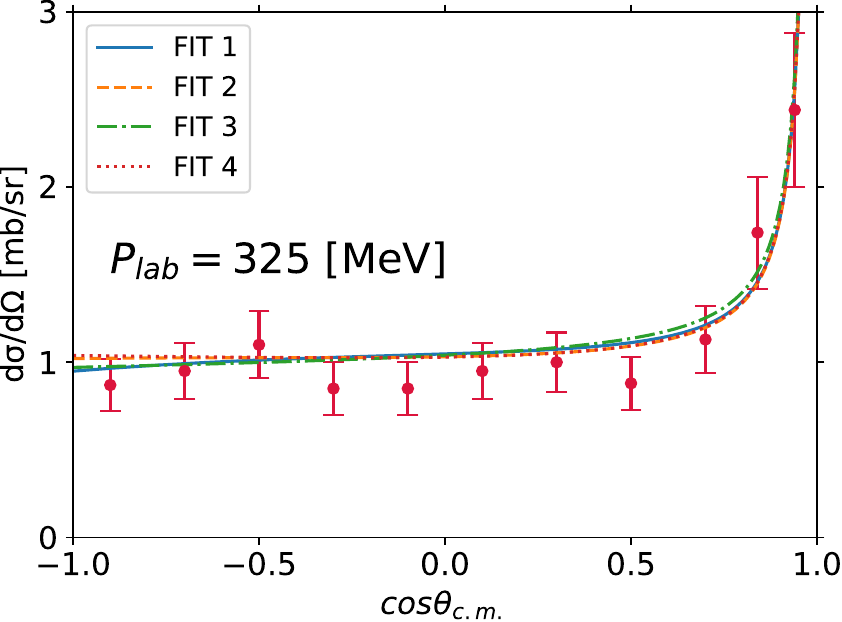}
    \end{minipage} &

    \begin{minipage}[c]{0.5\hsize}
      \centering
      \includegraphics[width=0.65\textwidth]{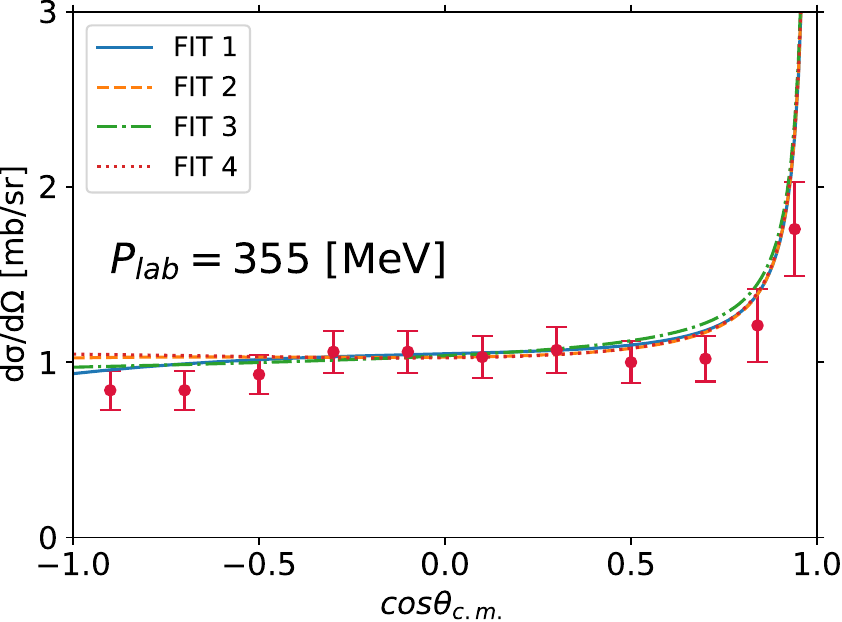}
    \end{minipage} \\

    \begin{minipage}[c]{0.5\hsize}
      \centering
      \includegraphics[width=0.65\textwidth]{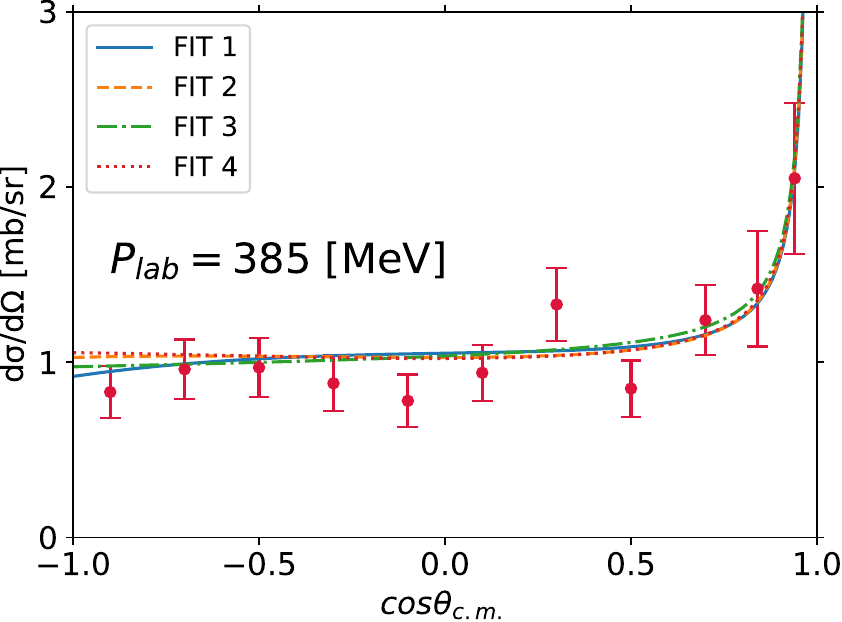}
      .        \end{minipage} &

    \begin{minipage}[c]{0.5\hsize}
      \centering
      \includegraphics[width=0.65\textwidth]{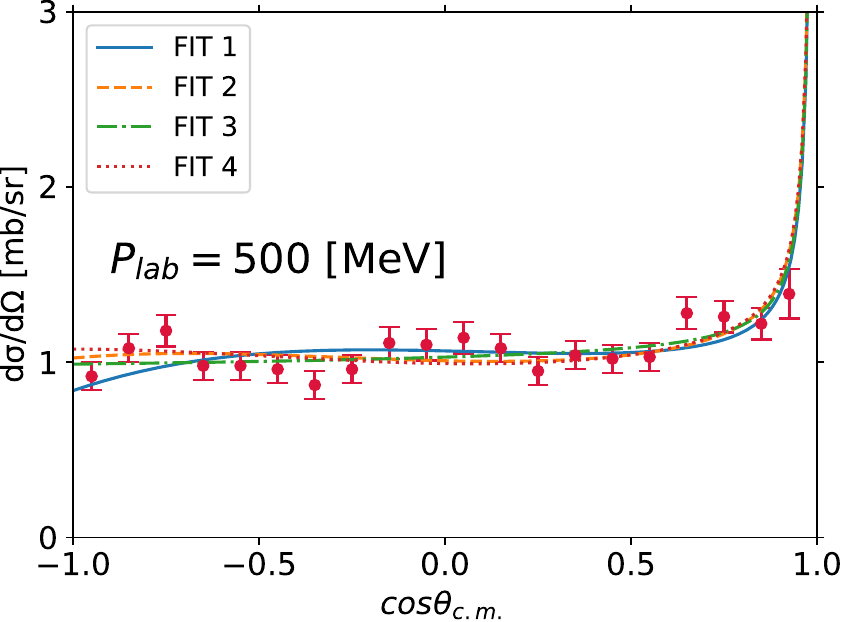}
    \end{minipage} \\

    \begin{minipage}[c]{0.5\hsize}
      \centering
      \includegraphics[width=0.65\textwidth]{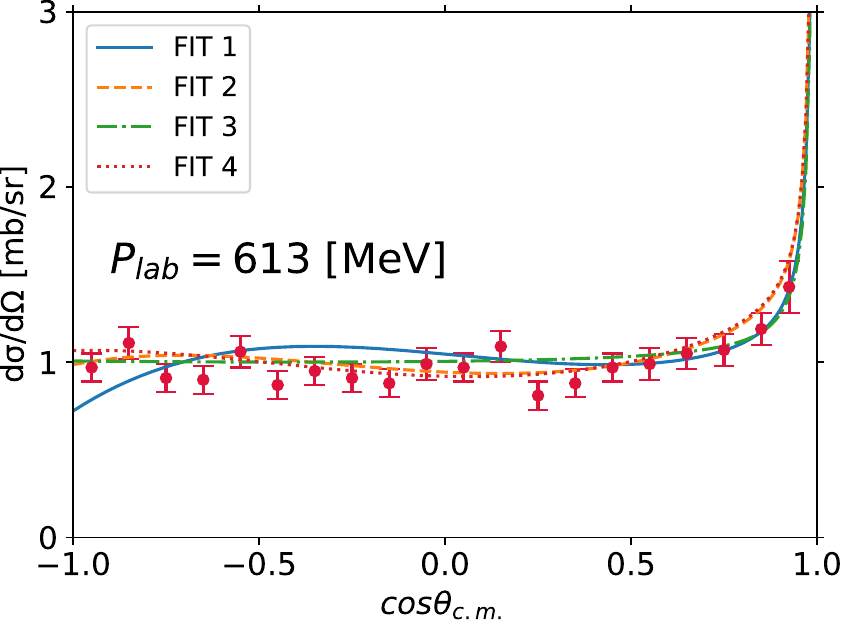}
    \end{minipage} &

    \begin{minipage}[c]{0.5\hsize}
      \centering
      \includegraphics[width=0.65\textwidth]{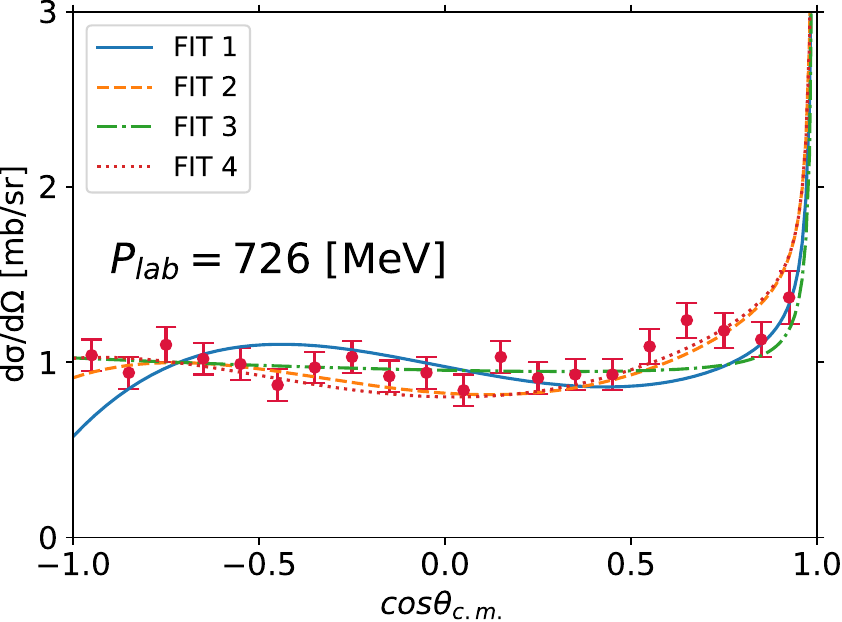}
    \end{minipage}
  \end{tabular}
  \caption{Calculated differential cross sections of the $K^{+}p$ elastic scattering in comparison with the experimental data of Ref. \cite{Cameron:1974xx}.}
  \label{fig:dCS_Kp}
\end{figure}

% In Fig.~\ref{fig:crosssection_I_1}, we show the $I=1$ total cross section calculated with the determined LECs and compare with the experimental data.
% In Fig.~\ref{fig:dCS_Kp}, we show the $K^+ p$ differential cross sections calculated with the determined LECs up to $P_\mathrm{lab} = 800\MeV/c$. The figure shows that our constructed scattering amplitudes reproduce the experimental data very well. Therefore chiral perturbation theory works well in the energy region we consider. The behavior of these calculated cross sections is very similar. This is evident in \Cref{tab:LECs}, because the values of the four sets of the $I=1$ LECs are relatively close; FIT~3 is a little further away than the other three, but still the signs of the LECs are consistent and not distant for all four.
In \Cref{fig:crosssection_I_1,fig:dCS_Kp}, we show our numerical results for the $I=1$ total cross section and the $K^+ p$ elastic differential cross sections calculated with the determined LECs, respectively, and compare them with the experimental observations. For the total cross section in \Cref{fig:crosssection_I_1} we use the scattering amplitude calculated only with the strong interaction, while the $K^+ p$ differential cross sections in \Cref{fig:dCS_Kp} include the Coulomb correlations formulated in \Cref{Sec:Coulomb}. In both figures, four sets of the determined LECs reproduce the experimental observations very well in the same manner. It is notable that chiral perturbation theory works well to reproduce the $I=1$ $K N$ amplitude in the energy region that we consider.
Some deviations among four fittings get evident from $P_\mathrm{lab} = 500 \MeV/c$ in the $K^+ p$ differential cross section.

\begin{figure}[H]
  \centering
  \includegraphics[width=0.6\textwidth]{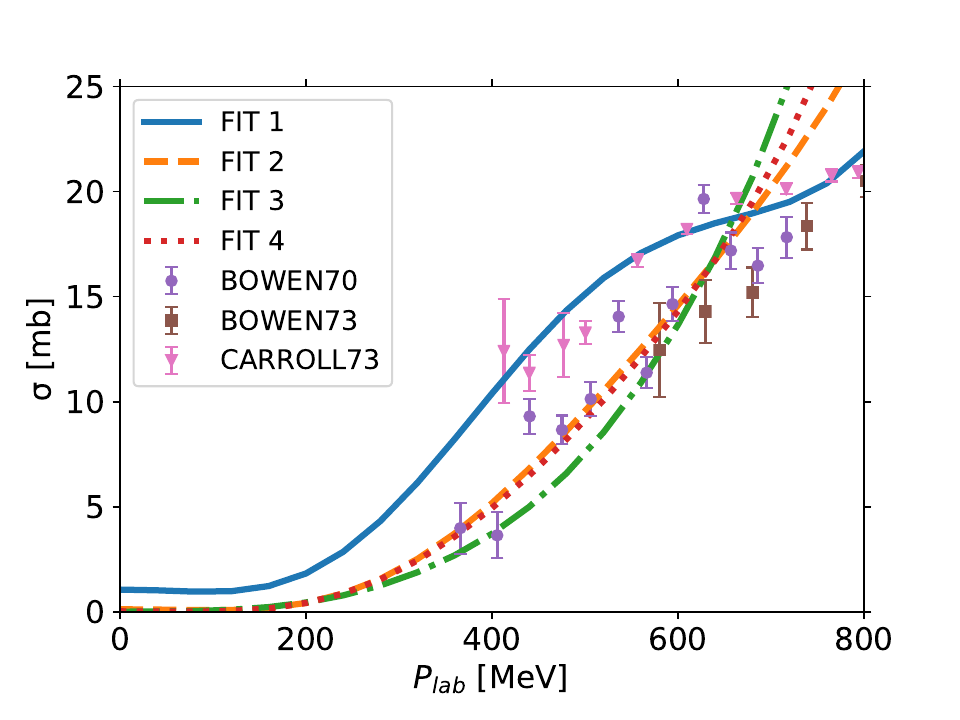}
  \caption{$I=0$ $K^+ N$ total cross sections calculated with the determined LECs given in Table.~\ref{tab:LECs} in comparison with the experimental data \cite{Carroll:1973ux,Bowen:1970azd,Bowen:1973ga}.
  }
  \label{fig:crosssection_I_0}
\end{figure}

\begin{figure}[H]
  \begin{tabular}{cc}
    \begin{minipage}[c]{0.5\hsize}
      \centering
      \includegraphics[width=0.80\textwidth]{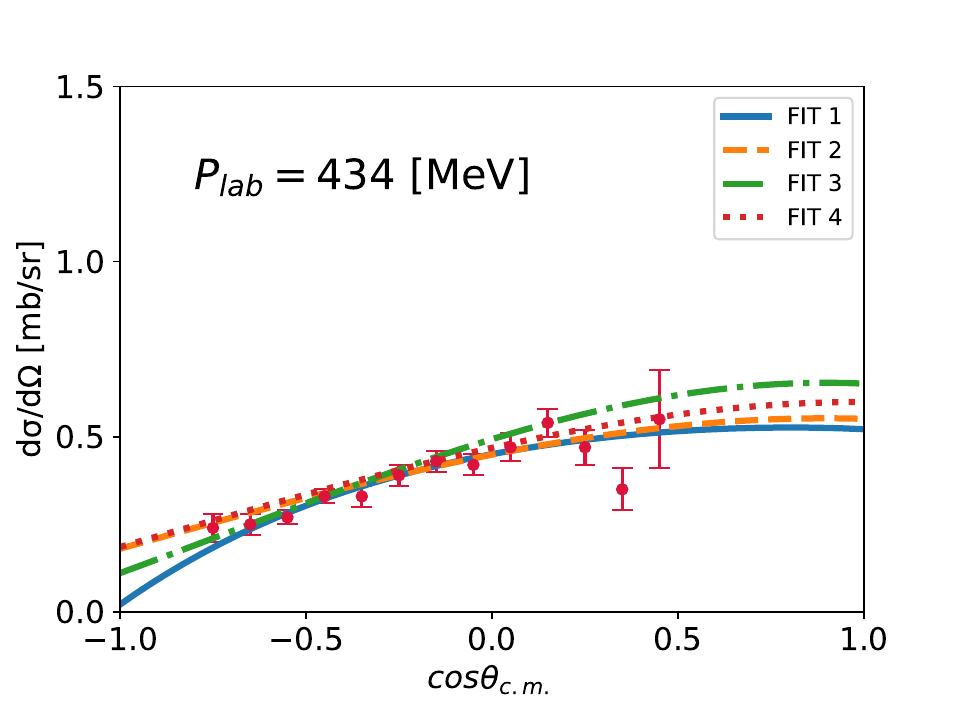}
    \end{minipage} &

    \begin{minipage}[c]{0.5\hsize}
      \centering
      \includegraphics[width=0.80\textwidth]{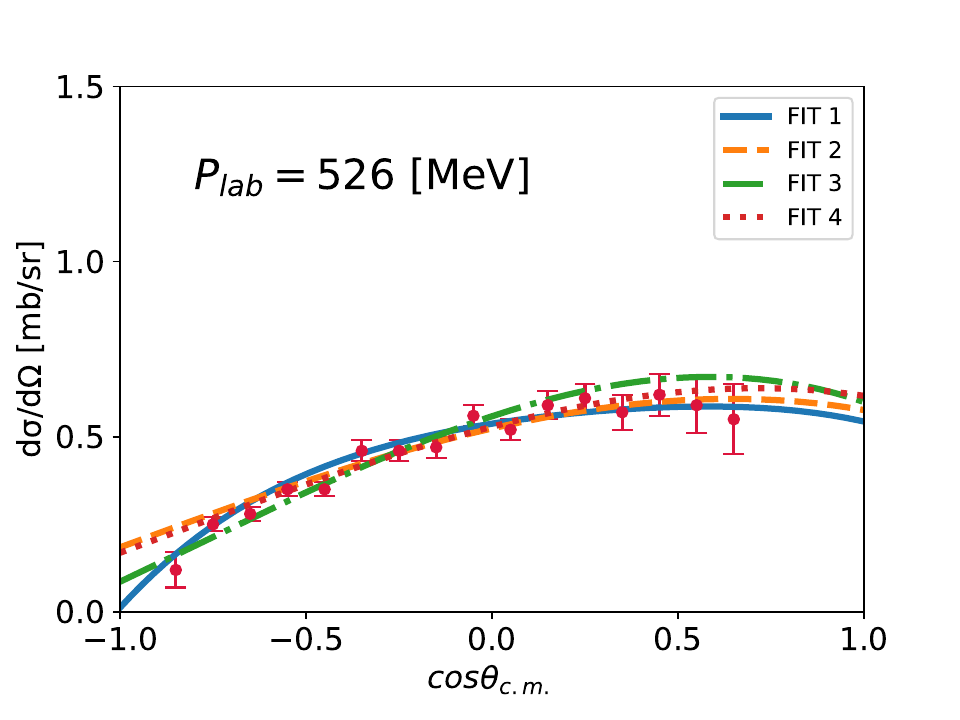}
    \end{minipage} \\

    \begin{minipage}[c]{0.5\hsize}
      \centering
      \includegraphics[width=0.80\textwidth]{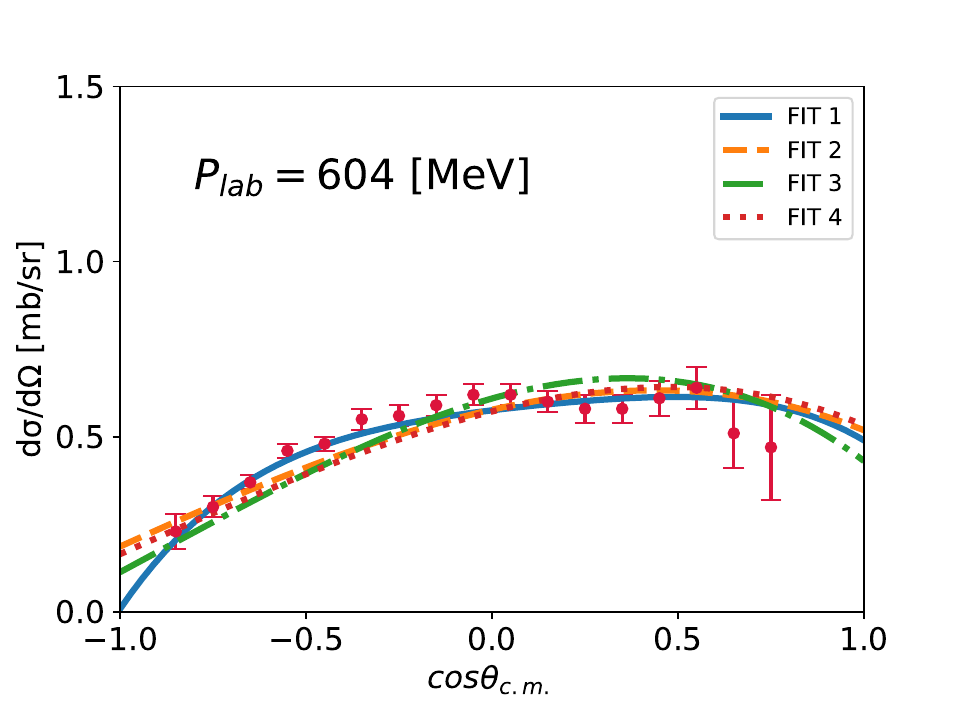}
    \end{minipage} &

    \begin{minipage}[c]{0.5\hsize}
      \centering
      \includegraphics[width=0.80\textwidth]{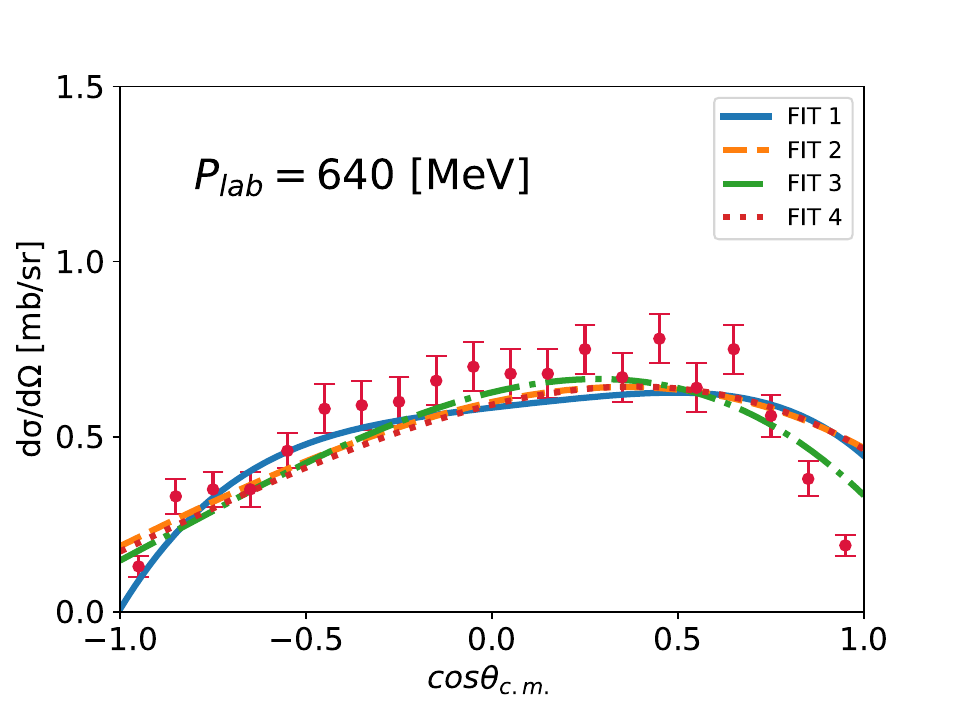}
    \end{minipage} \\
    \begin{minipage}[c]{0.5\hsize}
      \centering
      \includegraphics[width=0.80\textwidth]{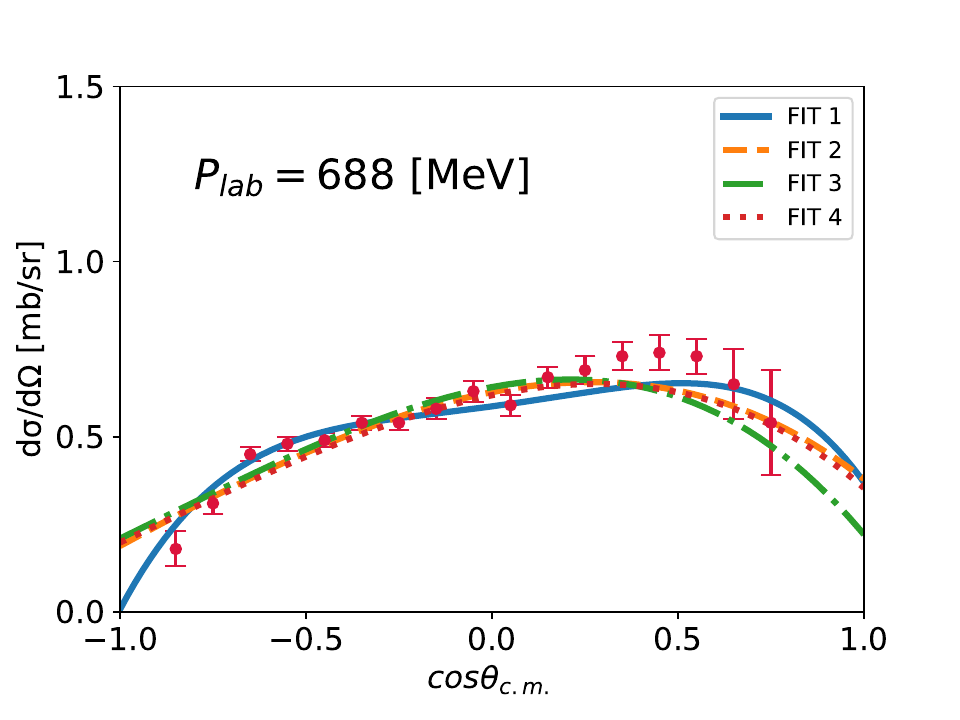}
    \end{minipage} &

    \begin{minipage}[c]{0.5\hsize}
      \centering
      \includegraphics[width=0.80\textwidth]{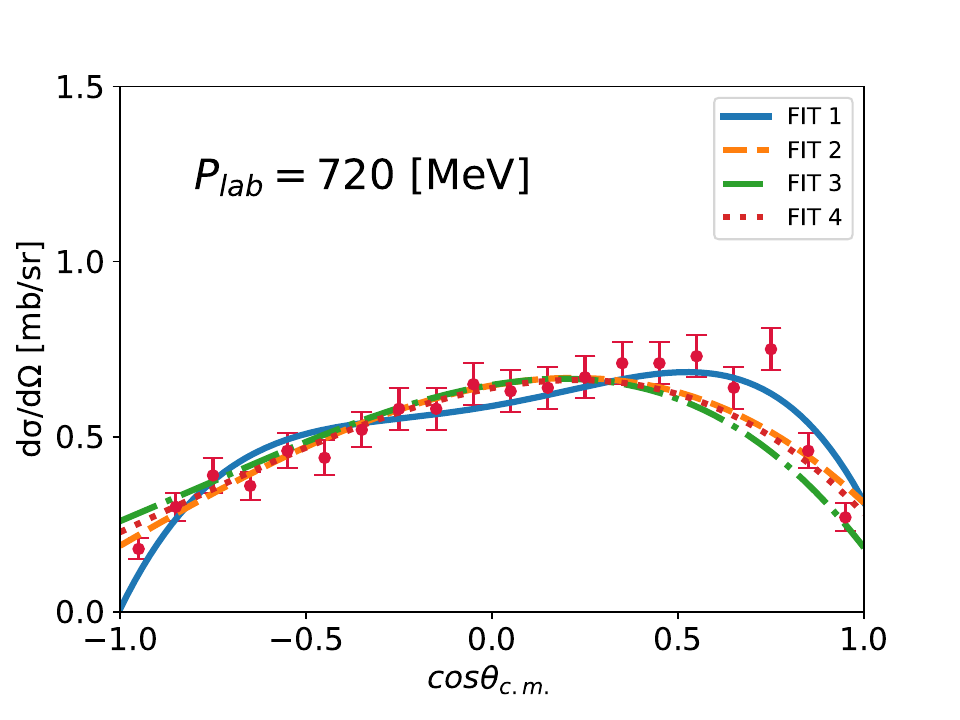}
    \end{minipage} \\

    \begin{minipage}[c]{0.5\hsize}
      \centering
      \includegraphics[width=0.80\textwidth]{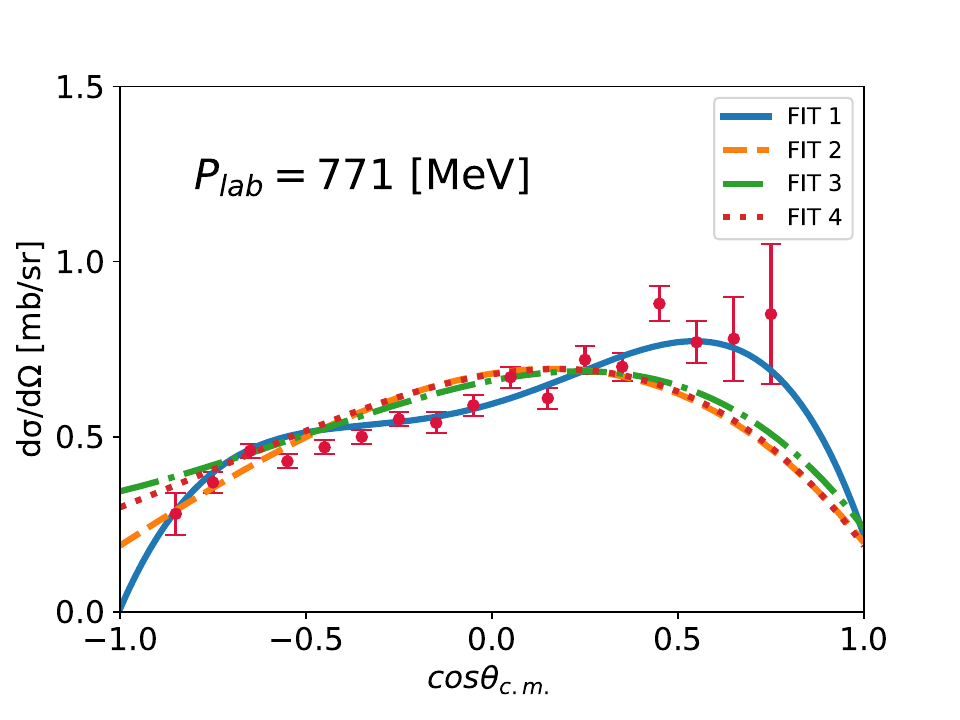}
    \end{minipage} &

    \begin{minipage}[c]{0.5\hsize}
      \centering
      \includegraphics[width=0.80\textwidth]{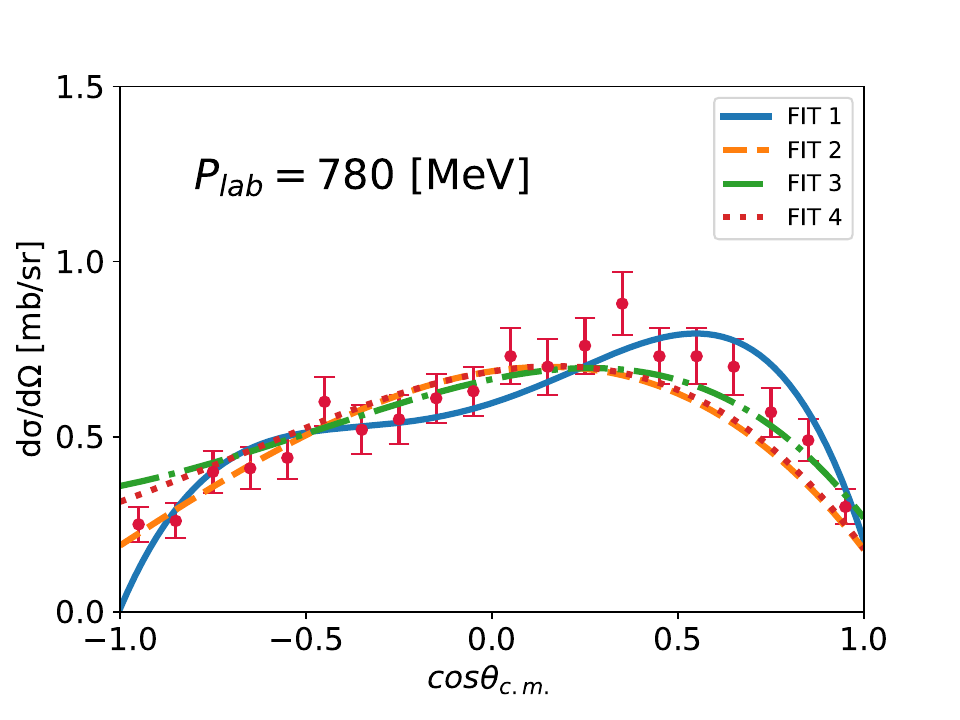}
    \end{minipage}
  \end{tabular}
  \caption{Calculated differential cross sections of $K^+ n\to K^0 p$ charge exchange scattering in comparison with the experimental data of Ref.~\cite{BGRT:1973bda,Damerell:1975kw}. The data for the momenta at the $P_\mathrm{lab} = 640, 720$ and $780 \MeV/c$ are taken from Ref.~\cite{BGRT:1973bda}, while the others are  from Ref.~\cite{Damerell:1975kw}.}
  \label{fig:dCS_Kn2Kp}
\end{figure}

% In Fig.~\ref{fig:crosssection_I_0}, we show the $I=0$ total cross section calculated with the determined parameters for each case and compare with the experimental data, and the partial wave contribution of the $I=0$ total cross section is seen in \Cref{fig:crosssection_I_0_breakdown}. Figure~\ref{fig:dCS_Kn2Kp} show the $K^+ n\to K^0 p$ charge exchange differential cross section.
% Each FIT reproduces the experimental data well. However the behavior of the calculated cross sections at low energies below $300\MeV$ is not determined because the experimental data do not exist in this energy region.
% This fact greatly affects the fitting of LECs. In fact, as seen in \Cref{tab:LECs}, the value of LECs changes depending on the presence or absence of resonance state and the choice of the experimental data as the $I=0$ total cross section, and the variation of $b^{I=0}$ is greater than that of $b^{I=1}$.
% In order to determine the value of LECs independent of the presence of the resonance and independent of the choice of experimental values for the $I=0$ total cross section, we need the experimental data with the small error for energies lower than the ones we have.
In \Cref{fig:crosssection_I_0,fig:dCS_Kn2Kp}, we show the $I = 0$ total cross section and the differential cross sections for the charge exchange process $K^+n \to K^0 p$ calculated with the determined LECs for each case, and we compare them with the experimental data. As stated above, for FIT~1 we use Carroll~1973 for the data of the $I=0$ total cross section, while in FITs~2, 3 and 4 Bowen~1970 is used. Each fit reproduces the experimental data well.
In particular, \Cref{fig:dCS_Kn2Kp} shows that these four fits reproduce the experimental data well up to $P_\mathrm{lab} = 720 \MeV/c$.  Nevertheless, it should be emphasized that we find some deviations among the fits in the total cross sections in low energies below $300 \MeV/c$. This is because the LECs are not constrained so much in low energies due to the lack of experimental data. In fact, as seen in \Cref{tab:LECs}, the values of LECs for $I=0$ are different in the fits. To fix the low-energy behavior of the scattering amplitude with $I=0$, experimental data below $300 \MeV/c$ are extremely important. It is also interesting to mention that the total cross sections obtained by FIT~2 and FIT~4 are almost the same up to $600 \MeV/c$. In these fits, we use the same experimental data (Bowen~1970) but FIT~4 includes the $P_{03}$ resonance contribution explicitly. Thus, our finding that FIT~2 and FIT~4 give a consistent result implies that the contribution of the $P_{03}$ resonance can be absorbed into the LECs as discussed in Ref.~\cite{Ecker:1988te}.
This situation can be understood by the fact that the obtained LECs for FIT~2 and FIT~4 are also almost equivalent but there is small deviation in the LECs for $I=0$.
These differences in the LECs represent the contribution of the $P_{03}$ resonance.

\begin{figure}[H]
  \begin{tabular}{cc}
    \begin{minipage}[c]{0.5\hsize}
      \centering
      \includegraphics[width=1.0\textwidth]{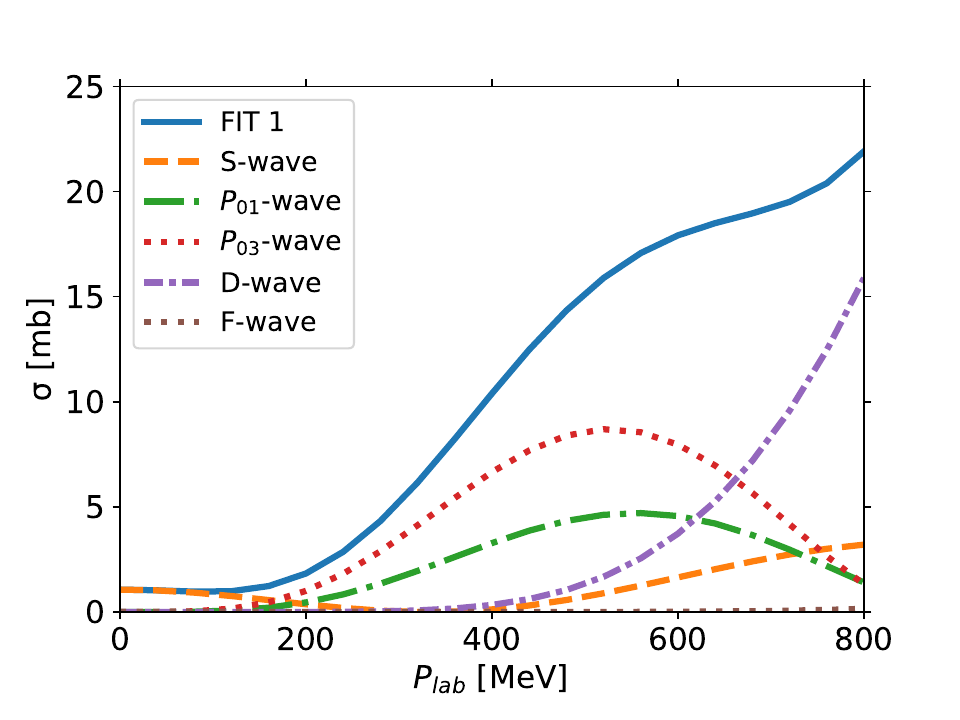}
    \end{minipage} &

    \begin{minipage}[c]{0.5\hsize}
      \centering
      \includegraphics[width=1.0\textwidth]{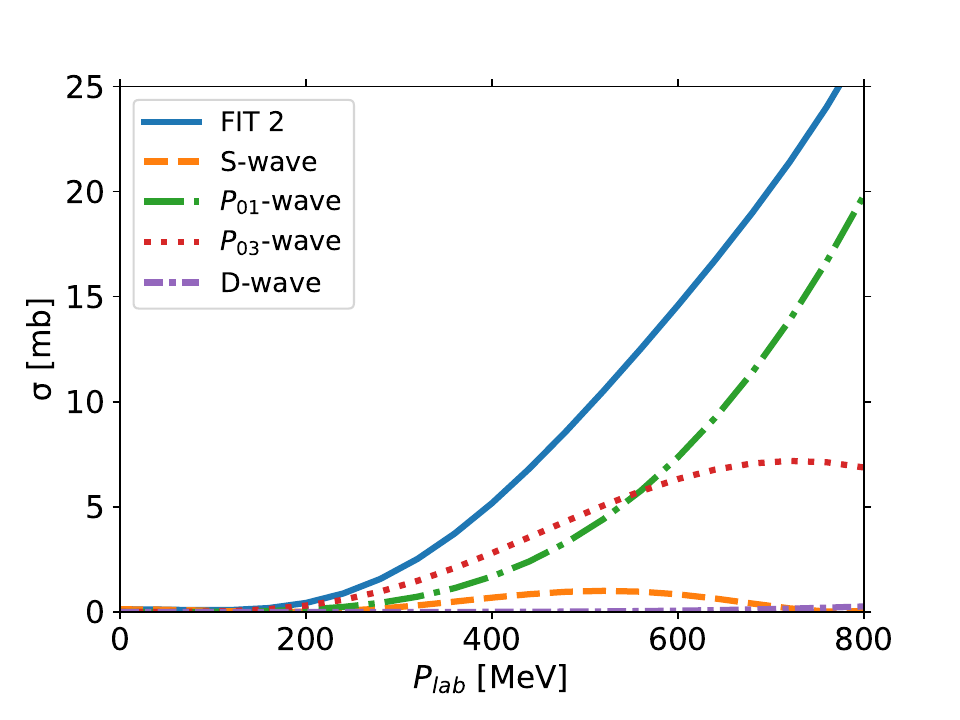}
    \end{minipage} \\

    \begin{minipage}[c]{0.5\hsize}
      \centering
      \includegraphics[width=1.0\textwidth]{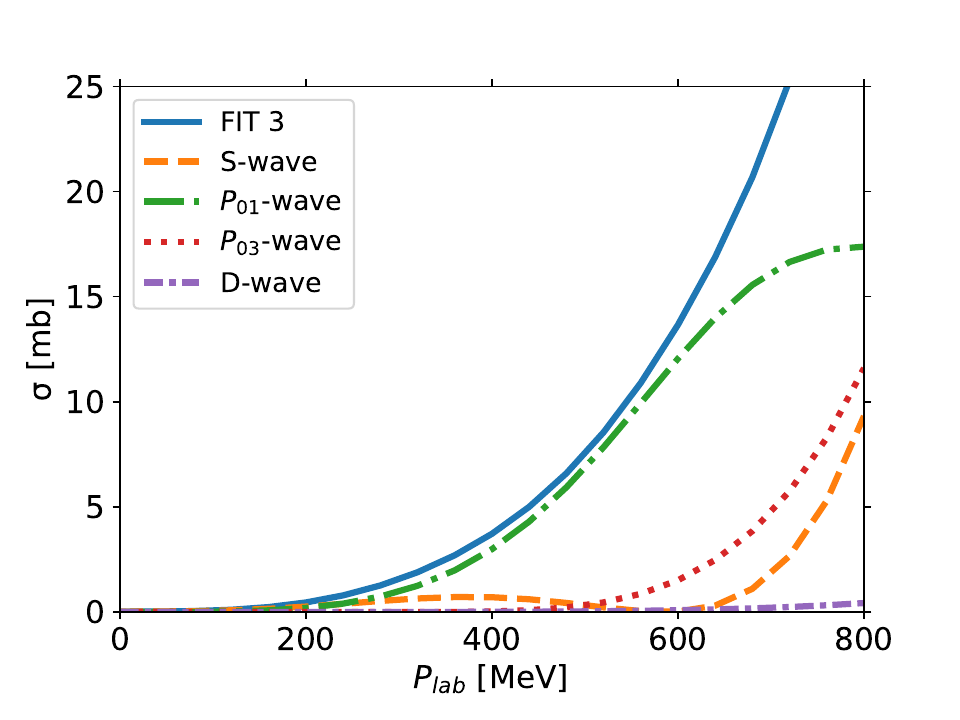}
    \end{minipage} &

    \begin{minipage}[c]{0.5\hsize}
      \centering
      \includegraphics[width=1.0\textwidth]{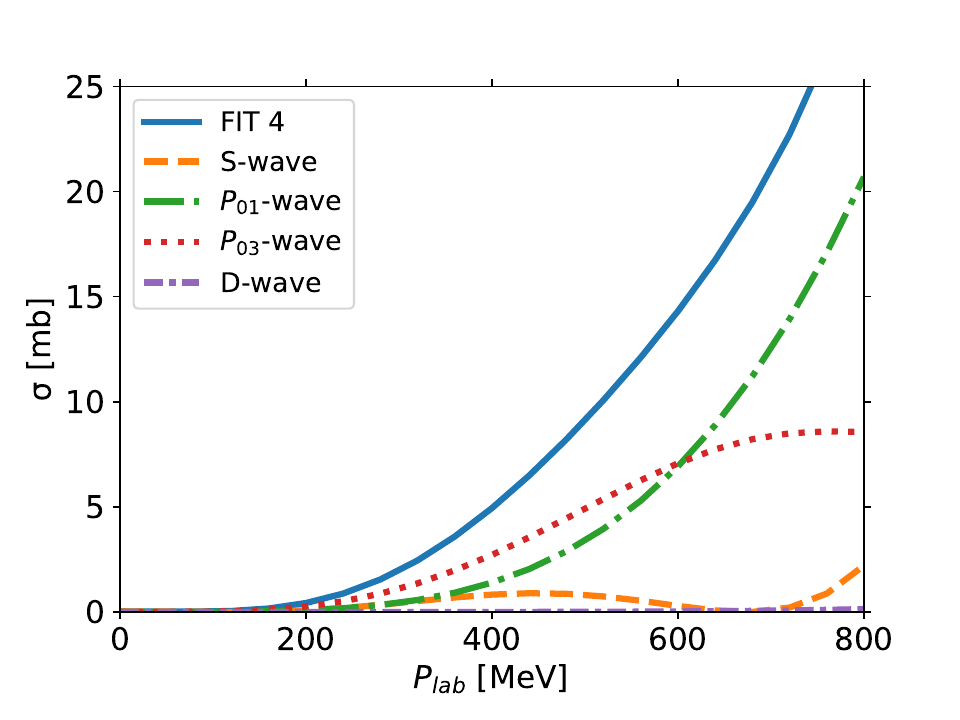}
    \end{minipage}
  \end{tabular}
  \caption{Partial wave contributions of $I=0$ $K^+ N$ total cross section calculated with the determined LECs.}
  \label{fig:crosssection_I_0_breakdown}
\end{figure}

% As seen in \Cref{fig:crosssection_I_0_breakdown}, the contribution of partial waves varies greatly with each FIT in the $I=0$ total cross section.
% In all fits except FIT~1, the influence of the $D$-wave is very small, while in FIT~1, the contribution of the $D$-wave is particularly large at high momentum.
% In FIT~2$\sim$4, $P$-wave is dominant. However, in FIT~3, the contribution of $P_{01}$-wave is larger since the $P_{01}$ resonance state is added to the $I=0$ amplitude; in FIT~4, the contribution of $P_{03}$-wave is larger than in FIT~3 since the $P_{03}$ resonance state is added to the $I=0$ amplitude, but interestingly, the behavior of the cross sections and their partial wave component calculated with FIT~2 and FIT~4 is so close that they are difficult to distinguish.
In \Cref{fig:crosssection_I_0_breakdown}, we show the partial wave decomposition of the $I = 0$ total cross sections obtained by the four fitting procedures. As seen in the figure, each fit provides different contributions of the partial waves. In FITs 2, 3, and 4 the contribution of the $D$-wave is negligibly small. This shows that the partial wave decomposition works well up to the $D$-wave for these fits. In contrast, in FIT 1, the $D$-wave contribution is particularly large at higher momentum. Nevertheless, we find that the $F$-wave contribution is negligibly small in FIT 1 as shown in \Cref{fig:crosssection_I_0_breakdown}. This indicates again that the partial wave decomposition works well up to the $F$-wave in FIT 1. In FITs 2, 3 and 4, $P$-waves give essential contributions, while $S$-wave contribution is found to be minor in all the fits especially for low energies. In FIT 3, the contribution of the $P_{01}$ partial wave is large reflecting the explicit introduction of the resonance contribution into the amplitude. The partial wave decomposition of FITs 2 and 4 are also consistent each other. This tells us again that FITs 2 and 4 are almost equivalent.

\begin{figure}[H]
  \begin{tabular}{cc}
    \begin{minipage}[c]{0.5\hsize}
      \centering
      \includegraphics[width=0.80\textwidth]{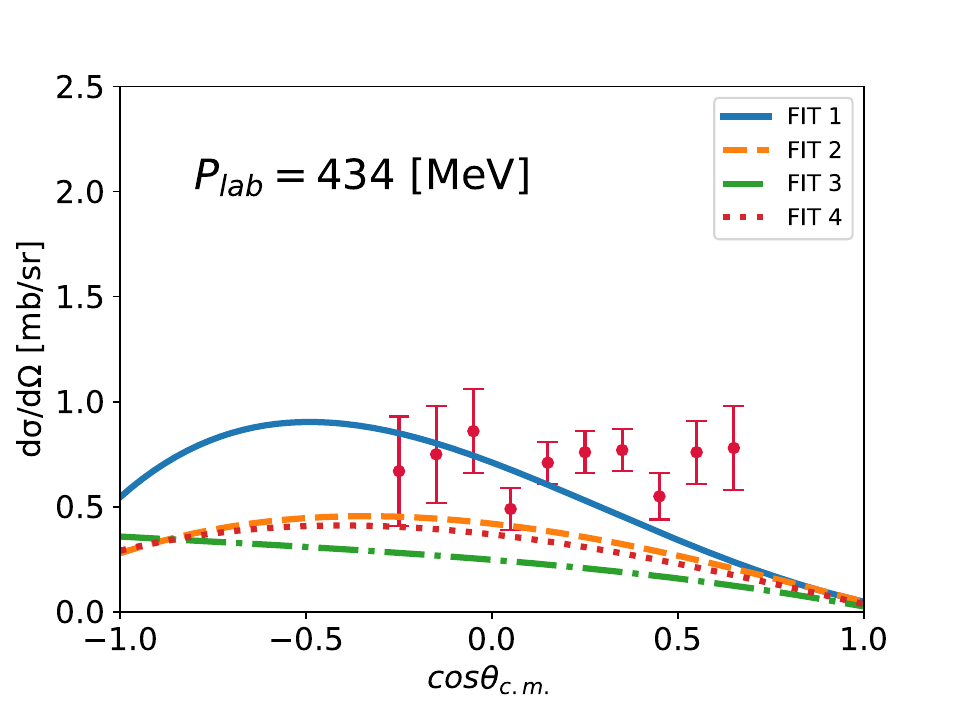}
    \end{minipage} &

    \begin{minipage}[c]{0.5\hsize}
      \centering
      \includegraphics[width=0.80\textwidth]{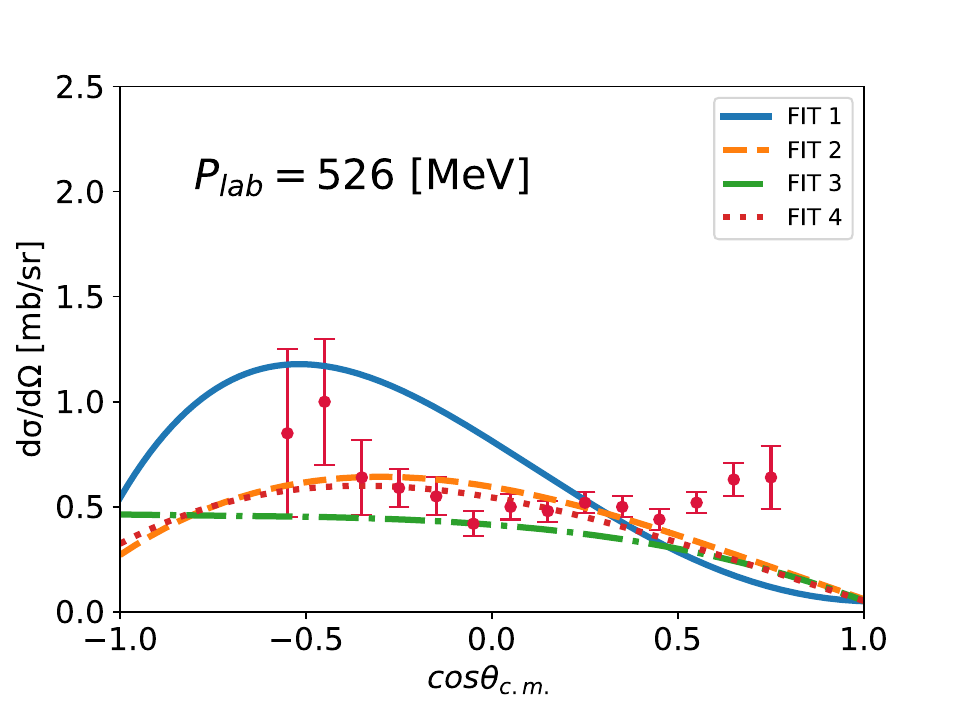}
    \end{minipage} \\

    \begin{minipage}[c]{0.5\hsize}
      \centering
      \includegraphics[width=0.80\textwidth]{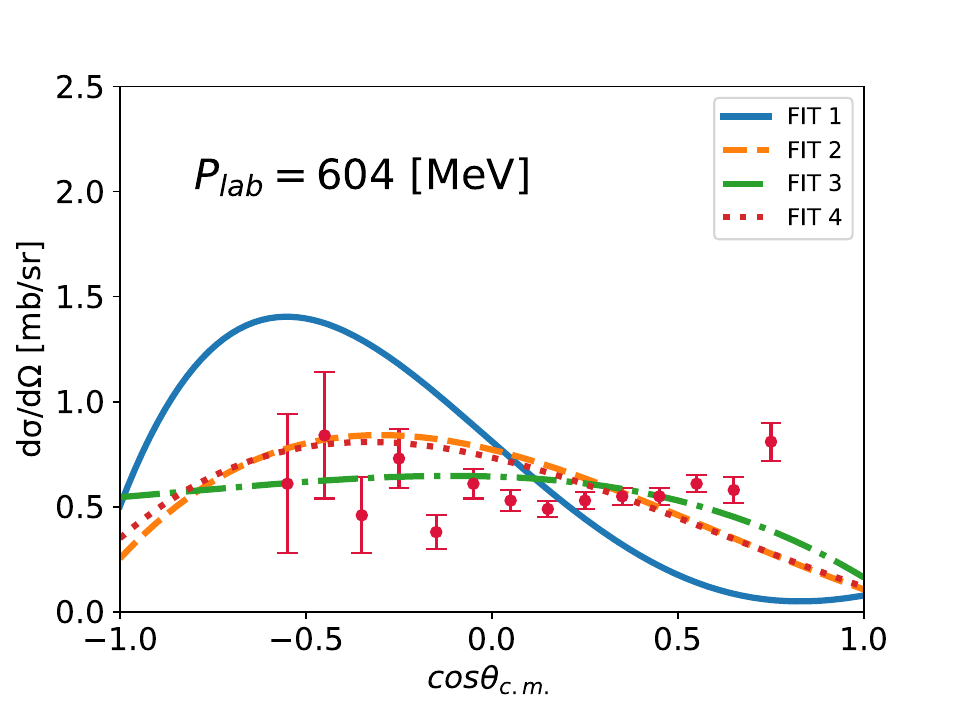}
    \end{minipage} &

    \begin{minipage}[c]{0.5\hsize}
      \centering
      \includegraphics[width=0.80\textwidth]{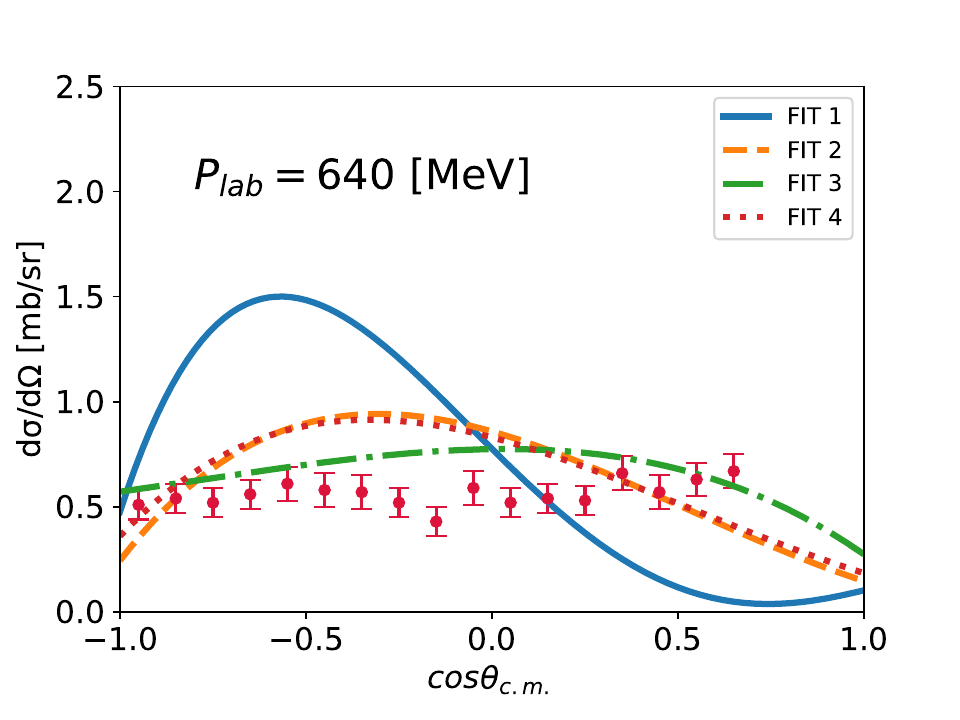}
    \end{minipage} \\
    \begin{minipage}[c]{0.5\hsize}
      \centering
      \includegraphics[width=0.80\textwidth]{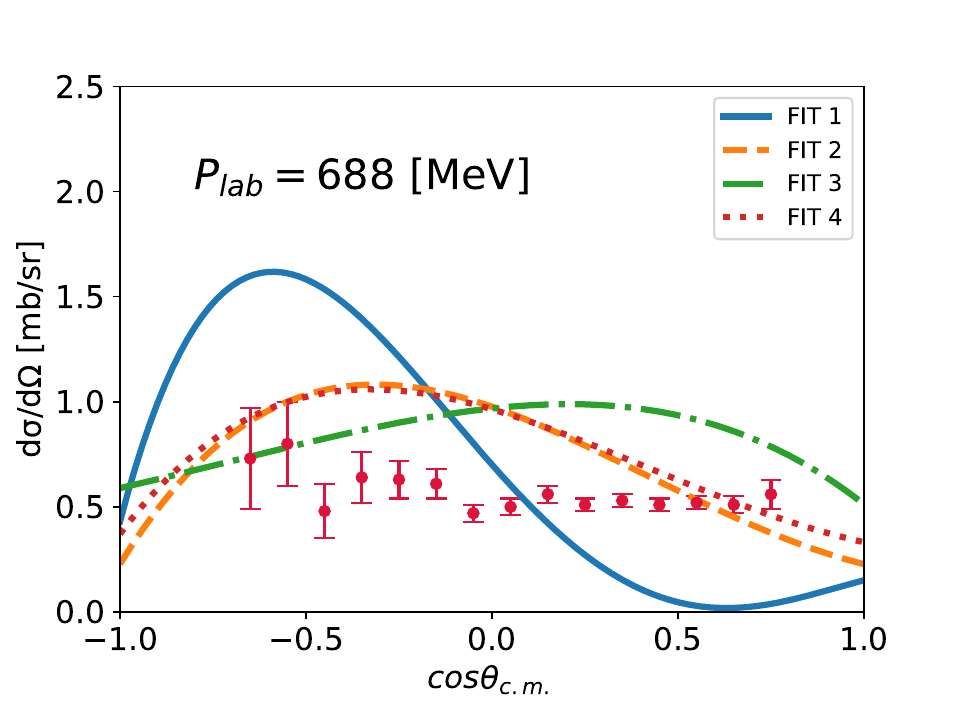}
    \end{minipage} &

    \begin{minipage}[c]{0.5\hsize}
      \centering
      \includegraphics[width=0.80\textwidth]{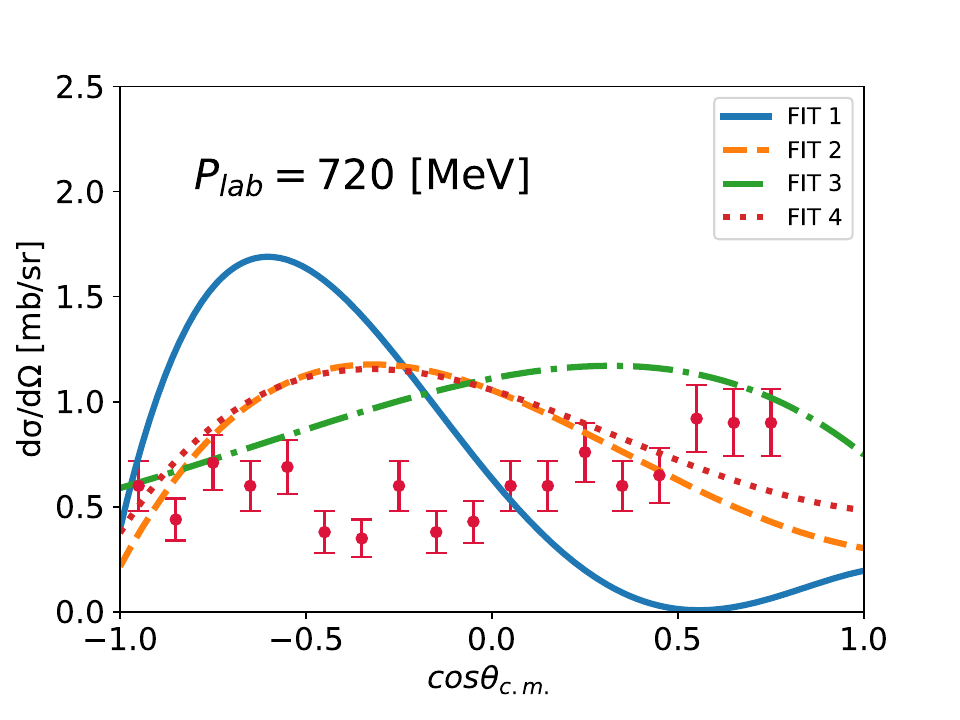}
    \end{minipage} \\

    \begin{minipage}[c]{0.5\hsize}
      \centering
      \includegraphics[width=0.80\textwidth]{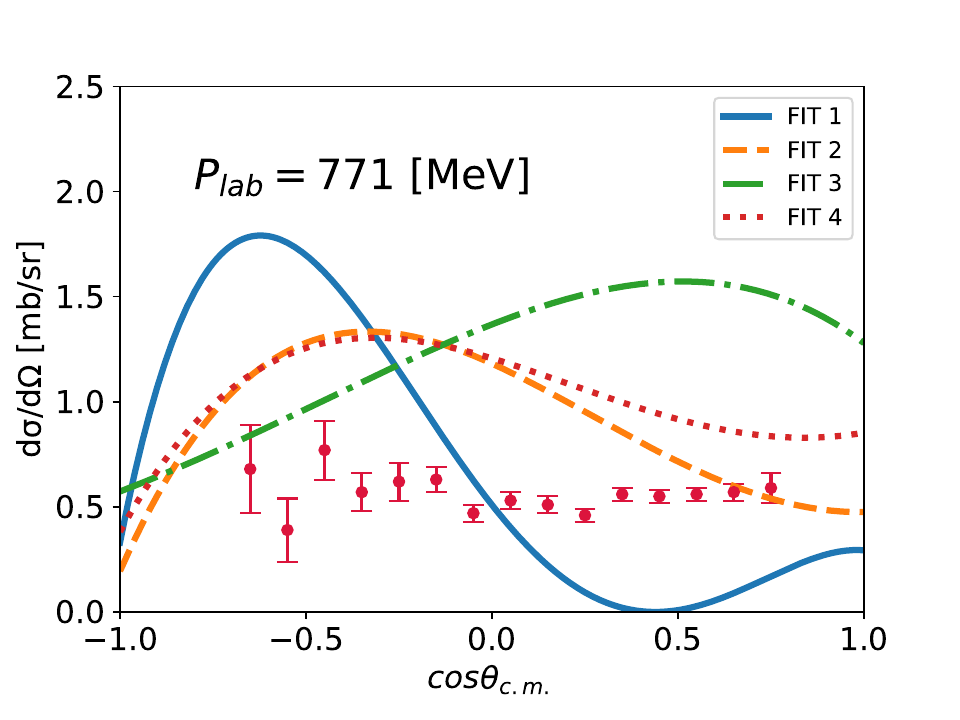}
    \end{minipage} &

    \begin{minipage}[c]{0.5\hsize}
      \centering
      \includegraphics[width=0.80\textwidth]{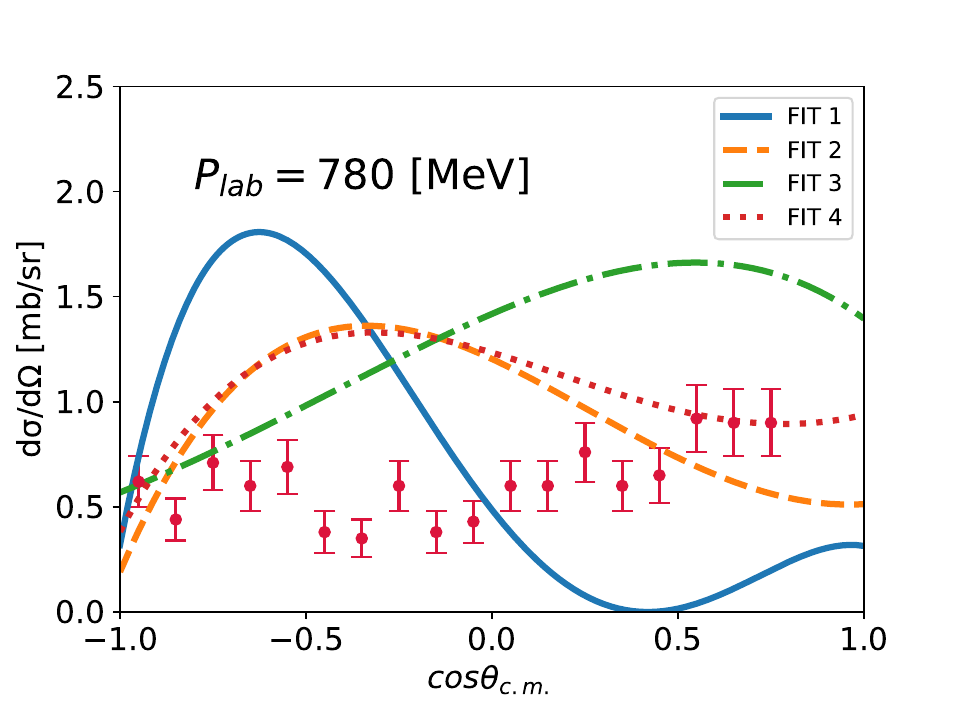}
    \end{minipage}
  \end{tabular}
  \caption{Calculated differential cross sections of $K^+ n$ elastic scattering in comparison with the experimental data of Ref.~\cite{BGRT:1973bda,Damerell:1975kw}. The data for the momenta at the $P_\mathrm{lab} = 640, 720$ and $780 \MeV/c$ are taken from Ref.~\cite{BGRT:1973bda}, while the others are from Ref.~\cite{Damerell:1975kw}.}
  \label{fig:dCS_Kn}
\end{figure}

% The differential cross sections for $K^+ n$ elastic scatterings are shown in \Cref{fig:dCS_Kn}. We do not use the experimental data of it for the fitting.
% From \Cref{fig:crosssection_I_1,fig:dCS_Kp,fig:crosssection_I_0,fig:dCS_Kn2Kp}, while the LECs are well fitted to the experimental data, Figure~\ref{fig:dCS_Kn} shows that they are very different with respect to the calculation of the $K^+ n$ elastic scatterings, and do not provide the results consistent with the experimental data. In addition, Figure~\ref{fig:dCS_Kn} also shows that the difference between FIT~2 and FIT~4 can be seen at $\cos\theta_\mathrm{c.m.} = 1$.
% Therefore, if accurate $K^+ n$ experimental data existed, they would be very restrictive with respect to the determination of the LECs and the presence or absence of resonance states.
In \Cref{fig:dCS_Kn}, we show our calculated results and the experimental data for the differential cross sections of the $K^+ n$ elastic scattering. Although the $K^+n$ elastic scattering data are not used for the fitting, the $K^+n$ elastic cross section should be reproduced according to the isospin symmetry, which is certainly good for hadronic reactions in these energies, because all of the theoretical calculations reproduce the cross sections of the $K^+p$ elastic and $K^+n\to K^0 p$ scatterings. Nevertheless, the experimental data are poorly reproduced in low energies and, especially, for higher energies the theoretical predictions are scattered among the fittings. Figure \ref{fig:dCS_Kn} also shows that the difference between FIT~2 and FIT~4 can be seen at $\cos \theta_\mathrm{c.m.} = 1$ for $P_\mathrm{lab} > 680 \MeV/c$, where the resonance contribution may be significant. This implies that forward scattering data for $P_\mathrm{lab} > 680 \MeV/c$ may give us important constraints on the wide resonance with $S = +1$.

\begin{table}[H]
  \centering
  \caption{Determined low-energy constants.
    %The values of the parameters except $w^{I}$ are shown in unit of $\GeV^{-1}$, and those of $w^{I}$s are shown in units of $\GeV^{-2}$.
    FIT~1 uses Carroll~1973~\cite{Carroll:1973ux} as the $I=0$ total cross section, while FIT~2 employs Bowen~1970 \cite{Bowen:1970azd}. Both cases do not introduce the broad resonance into the $I=0$ amplitude.
    FIT~3 considers the $P_{01}$ resonance by adding the resonance contribution, while FIT~4 takes account of the $P_{03}$ resonance. In FIT~3 and FIT~4, Bowen~1970 is used for the $I=0$ total cross section.}
  \begin{tabular}{cc|rrrr}
    \hline
     LEC & unit & \multicolumn{1}{c}{FIT~1} &
    \multicolumn{1}{c}{FIT~2} &
    \multicolumn{1}{c}{FIT~3} &
    \multicolumn{1}{c}{FIT~4}                                                                              \\\hline  
    $b^{I=1}$ & $[\GeV^{-1}]$  & $-1.07\pm 0.11$  & $-1.10\pm 0.10$ & $-0.11\pm 0.12$ & $-1.08\pm 0.11$ \\
    $d^{I=1}$ & $[\GeV^{-1}]$  & $-2.05\pm 0.20$  & $-2.00\pm 0.17$ & $-0.19\pm 0.19$  & $-1.97\pm 0.17$ \\
    $g^{I=1}$ & $[\GeV^{-1}]$  & $-0.82\pm 0.22$  & $-0.93\pm 0.18$ & $-0.80\pm 0.20$ & $-1.01\pm 0.19$ \\
    $h^{I=1}$ & $[\GeV^{-1}]$  & $3.67\pm 0.50$   & $4.07\pm 0.60$  & $0.91\pm 0.54$  & $4.21\pm 0.60$  \\
    $w^{I=1}$ & $[\GeV^{-2}]$  & $-0.76\pm 0.11$  & $-1.00\pm 0.10$ & $-0.36\pm 0.10$ & $-1.05\pm 0.10$ \\
    \hline
    $b^{I=0}$ & $[\GeV^{-1}]$  & $-3.66\pm 0.30$  & $1.45\pm 0.40$  & $2.36\pm 0.48$  & $2.29\pm 0.40$  \\
    $d^{I=0}$ & $[\GeV^{-1}]$  & $-9.21\pm 0.40$  & $-0.20\pm 0.40$ & $-1.42\pm 0.58$ & $-0.63\pm 0.50$ \\
    $g^{I=0}$ & $[\GeV^{-1}]$  & $1.46\pm 0.50$   & $6.10\pm 0.70$  & $8.27\pm 0.95$  & $8.07\pm 0.80$  \\
    $h^{I=0}$ & $[\GeV^{-1}]$  & $16.29\pm 0.70$  & $-3.99\pm 0.80$ & $-1.64\pm 0.96$ & $-4.91\pm 0.80$ \\
    $w^{I=0}$ & $[\GeV^{-2}]$  & $-0.57\pm 0.29$  & $4.23\pm 0.35$  & $4.92\pm 0.46$  & $4.99\pm 0.40$  \\
    \hline
    $v_-$ & $[\GeV^{-1}]$      & $42.89\pm 1.70$  & $12.32\pm 1.70$ & $5.00\pm 0.19$  & $10.12\pm 1.70 $ \\
    $v_+$ & $[\GeV^{-1}]$      & $-7.55\pm 0.90$  & $4.28\pm 0.90$  & $-3.63\pm 0.93$ & $4.74\pm 0.90$  \\
    \hline\hline
    $\chi^2_\mathrm{dof}$ &     & 2.41            & 2.74          & 2.95          & 2.96          \\
    \hline
  \end{tabular}
  \label{tab:LECs}
\end{table}
% \end{landscape}

\subsection{Behavior of in-medium quark condensate with strange quarks}
\label{sec:behavior}
In this section, we discuss the behavior of the in-medium quark condensate with strange quarks by using Eq.~(\ref{eq:uu_plus_ss_cond}) with the determined LECs in the previous section. It should be noted that we focus on the qualitative behavior of the quark condensate in the nuclear matter, because the condensate (\ref{eq:uu_plus_ss_cond}) is calculated under the linear density approximation. In addition, to separate out $\langle \bar ss \rangle^*$ from Eq.~(\ref{eq:uu_plus_ss_cond}) one needs to calculate the in-vacuum condensates with taking into account of the $\SUN{3}$ breaking effect. We also note that, as we have seen in the previous section that the LECs are not determined well with the existing data, the discussion on the detailed value of the in-medium quark condensate is not in the scope of this paper.

As seen in \Cref{eq:uu_plus_ss_cond}, the sign of the coefficient of the linear density, $3b^{I=1} + b^{I=0}$, determines whether the condensate increases or decreases in the nuclear matter. The slope parameters obtained in the present calculation are summarized in \Cref{tab:slope}. There the central values of the determined LECs are used. The table shows that the determined slope parameters are mostly negative, which means that the magnitude of the quark condensate decreases as the density increases, but their values differ in a wide range. For comparison, we also show the values of the slope parameters evaluated by the LECs determined in other calculations based on the baryon masses. As a theoretical calculation, we use the LECs determined by lattice calculation. Reference~\cite{Geng:2013xn} expressed the octet baryon masses in terms of the LECs by using an ${\cal O}(p^4)$ chiral perturbation theory in the extended-on-mass-shell scheme and determined the LECs by fitting them to lattice QCD calculation with various values of the quark masses. In addition, we also consider the LECs in more phenomenological determinations. The values of $b_F$ and $b_D$ can be fixed by the mass splitting of the octet baryons in the leading order of chiral perturbation theory as given, for instance, in Ref.~\cite{Kubis:2000aa,Holmberg:2018dtv}, while we fix the value of $b_0$ by the $\sigma_{\pi N}$ term together with $b_F$ and $b_D$ using the relation between the LECs of the SU(2) and SU(3) chiral perturbation theories given in Ref.~\cite{Hubsch:2021nih} as  
\begin{align}\label{eq:assumption_b0}
  2b_0 + b_D + b_F 
  = 2b_0 + b^{I=1} - b^{I=0} 
  = 2c_1
\end{align}
where $c_1$ is one of the SU(2) LECs and is given by $c_1 = - \sigma_{\pi N}/(4m_\pi^2)$ in the leading order of chiral perturbation theory. Its value can be fixed as $c_1 = -0.78\GeV^{-1}$ by using $\sigma_{\pi N} = 60$~MeV as suggested recently in Refs.~\cite{Alarcon:2011zs,Chen:2012nx,Hoferichter:2015dsa,Yao:2016vbz,RuizdeElvira:2017stg}. This value is also consistent with a recent analysis based on pionic atom data~\cite{Friedman:2019zhc}. With this value, however, the linear density approximation provides as larger as 50\% reduction of the quark condensate in magnitude at the saturation density, while a smaller value, $\sigma_{\pi N} \simeq 45 \MeV$, is preferable to reproduce 35\% reduction in the linear density analysis.
Anyway, it is a good advantage of the present work that the slope parameter is directly determined by the physical observables without using the value of the $\sigma_{\pi N}$ term. 
{\red In addition, we also compare the LECs obtained by a global fitting performed in Ref.~\cite{Lu:2018zof}. There the $\pi N$ and $KN$ scattering amplitudes were calculated using chiral perturbation theory up to ${\cal O}(p^3)$ for the $\pi N$ channel and ${\cal O}(p^2)$ for the $KN$ channel. (They also performed calculation with $KN$ amplitudes including one loop contributions, which are a part of ${\cal O}(p^3)$). The LECs were determined commonly by using $\pi N$ and $KN$ phase shift analyses. For the $KN$ scattering they used the SP92 solution \cite{Hyslop:1992cs} and took the $KN$ phase shifts only in low energies between $P_\mathrm{lab} = 25 \MeV/c$ to $257 \MeV/c$. Note that our study uses direct scattering data in much wider range up to 800 MeV/c, where the scattering data are available from $P_\mathrm{lab} = 145 \MeV/c$ for the $K^+p$ channel and from $434 \MeV/c$ for the $K^+n$ channel. In \Cref{tab:slope}, we show the LECs obtained by the fitting strategy one in Ref.~\cite{Lu:2018zof} where they did not consider the constraint on the LECs from the baryon masses.}

% We adopt the $\SUN{3}$ low-energy constants obtained by other calculations and summarized in \Cref{tab:SU3LECs} in order to see the dependency of the parameter choice.
% The first set was obtained by fitting to lattice QCD data of octet baryon masses \cite{Geng:2013xn}.
%% The difference between the first and second set lies in the number of masses that were considered for the fit. 
% The second set was obtained by Ref.~\cite{Kubis:2000aa} via fitting to experimental baryon octet data, this was also summarized in Ref.~\cite{Holmberg:2018dtv}.
% Since Ref.~\cite{Kubis:2000aa} only determined $b_D$ and $b_F$, we fix $b_0$ 
% by using the relation formula between SU(2) and SU(3) LECs \cite{Hubsch:2021nih}
% \begin{align}\label{eq:assumption_b0}
%  2b_0 + b_D + b_F = 2b_0 + b^{I=1} - b^{I=0} = 2c_1
% \end{align}
% where $c_1$ is one of the SU(2) LECs. Here we use $c_1 = -0.59\GeV^{-1}$ \cite{Gasser:1990ce}.

\begin{table}[tb]
  \centering
  \caption{Values of the slope parameter $(3b^{I=1} + b^{I=0})$ appearing in Eq.~(\ref{eq:uu_plus_ss_cond}) obtained in the present work. The central values of the determined LECs are used. FIT~3$'$ is a second best solution of the fitting procedure FIT3. The values of the slope parameters calculated with the LECs in other calculations, a theoretical calculation using lattice data~\cite{Geng:2013xn}, {\red a phenomenological calculation using the octet baryon masses~\cite{Kubis:2000aa,Holmberg:2018dtv} together with the $\sigma_{\pi N}$ term (see text) and a global fitting of LECs in chiral perturbation theory using $\pi N$ and $KN$ phase shift analyses~\cite{Lu:2018zof}, are also shown. These works are referred as Th., Pheno.\ and ChPT, respectively. The values of the relevant LECs for these calculations are $(b_0, b_D, b_F) = (-0.609,0.225,-0.404)$, $(-0.711,0.060,-0.190)$ and $(b^{I=0},b^{I=1})=(0.136,-0.270)$, respectively.} 
  %The values are shown in units of $\GeV^{-1}$.
  }
  \begin{tabular}{l|ccccc|ccc}
    \hline
        $[\GeV^{-1}]$  & FIT 1 & FIT 2 & FIT 3 & FIT 4 & FIT 3$^\prime$ & Th. & Pheno. & ChPT\\
    \hline
    $3b^{I=1} + b^{I=0}$ & $-6.87$ & $-1.86$ & $2.02$ & $-0.96$ & $-1.98$ & $-1.36$ & $-2.47$ & -0.674  \\                
    \hline
  \end{tabular}
%  \begin{tabular}{ll|ccccc|cc}
%    \hline
%        &  & FIT 1 & FIT 2 & FIT 3 & FIT 4 & FIT 3$^\prime$ & Th. & Pheno. \\
%    \hline
%    $3b^{I=1} + b^{I=0}$& $[\GeV^{-1}]$ & $-6.87$ & $-1.86$ & $2.02$ & $-0.96$ & $-1.98$ & $-1.36$ & $-2.47$   \\                
%    \hline
%  \end{tabular}
  \label{tab:slope}
\end{table}

\begin{figure}[H]
  \centering
  \includegraphics[width=0.6\textwidth]{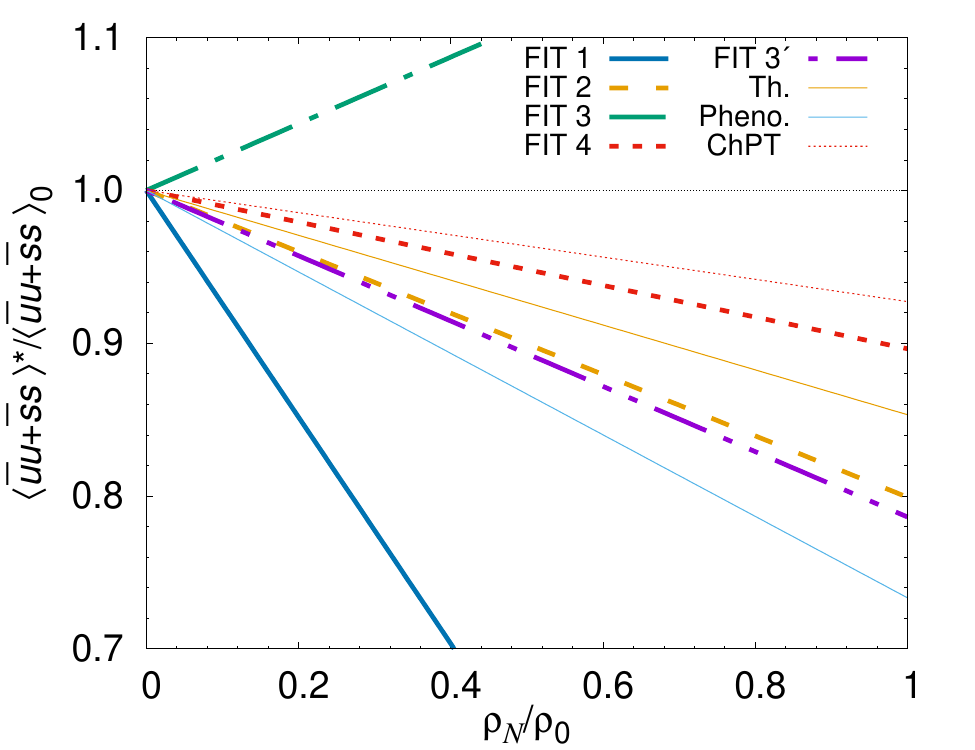}
%  {fig_all/plot_uu_plus_ss.pdf}
  \caption{Density dependence of the in-medium quark condensate with strange quarks normalized by the in-vacuum condensate calculated with the slope parameters given in \Cref{tab:slope}. We use $\rho_0 = 0.17$ fm$^{-3}$.}
  \label{fig:uu_plus_ss_cond}
\end{figure}

In \Cref{fig:uu_plus_ss_cond}, we show the density dependence of the in-medium quark condensate with strange quarks normalized by the in-vacuum condensate. The calculation is done with \Cref{eq:uu_plus_ss_cond} using the slop parameters shown in \Cref{tab:slope}. The behavior of the in-medium condensate is highly dependent on the choice of the parameter sets. The quark condensates with FITs~2 and 4 decrease in magnitude moderately as the density increases and the reduction at the saturation density $\rho_0$ is found to be about 10$\sim$20\%. The baryon mass determinations of the LECs also give consistent results. The quark condensate calculated with FIT~1 decreases significantly and reaches out of the range of reliability. This implies that the current status of the $K^+ N$ scattering data may not have enough quality for the determination of the LECs. 

In contrast to the findings with FITs~1, 2 and 4, the quark condensate calculated with FIT~3 largely increases in magnitude. This behavior might be unnatural in the context of the partial restoration of DB$\chi$S in finite density. For FIT~3, which uses Bowen 1970 for the $I=0$ total cross section and introduces the $P_{01}$ broad resonance, we find a second best solution that minimizes \Cref{eq:chi-squared}. This solution is named FIT~3$'$ and its LECs are shown in \Cref{tab:FIT3_2}. Comparing the LECs for $I=1$ of FIT~3$'$ with those of the other fits, we find that FIT~3$'$ has LECs closer to FITs~1, 2 and 4. The results of the calculations of the slope parameter and the in-medium quark condensate using the LEC of FIT~3$'$ are also shown in \Cref{tab:slope} and \Cref{fig:uu_plus_ss_cond}, respectively, which show that the density dependence of the quark condensate for FIT3$'$ is consistent with FITs~2 and 4. 
{\red The existence of a more reasonable solution with a similar $\chi^2_\mathrm{d.o.f.}$ value does not imply that the fitting procedure 3, where we have assumed a $P_{01}$ resonance, should be immediately ruled out.}
The fact that there is another independent solution to minimize $\chi^2_\mathrm{d.o.f.}$ with a similar value may indicate that the LECs giving the smallest value of $\chi^2_\mathrm{d.o.f.}$ can be changed with more experimental observations in the future, {\red such as $K^+d$ reaction at J-PARC~\cite{Ahn:2023hiu} and $K^0 p$ reaction at K-Long Facility in Jefferson Laboratory~\cite{KLF:2020gai}.}
% (see also Ref.~\cite{Amaryan:2024koq})

The choice of the experimental data of the $I=0$ total cross sections and the presence or absence of the resonance state in $I=0$ $K^+ N$ scattering have a significant impact on the determination of the LECs.
%which is important for the behavior of the condensate.
Therefore, we emphasize that, in order to determine the behavior of the in-medium quark condensate with strange quarks more precisely, it is extremely important to determine experimental values accurately and consistently with isospin symmetry at a wide range of energy in particular much lower than $P_\mathrm{lab} = 600\MeV/c$ where the effect of the resonance state are less significant.

% \subsection{Discussion}\label{sec:discussion}
% \subsection{Another solution with the $P_{01}$ resonance}\label{sec:anotherFIT3}

\begin{table}[H]
  \centering
  \caption{Same as FIT~3 in \Cref{tab:LECs} but the LECs for a second best solution.}
  \begin{tabular}{cc|r}
   \hline
    LEC & unit & \multicolumn{1}{c}{FIT~3$'$} \\\hline 
    $b^{I=1}$   & $[\GeV^{-1}]$   & $-0.39\pm 0.12$     \\
    $d^{I=1}$   & $[\GeV^{-1}]$   & $-0.69\pm 0.18 $     \\
    $g^{I=1}$   & $[\GeV^{-1}]$   & $-1.07\pm 0.21$     \\
    $h^{I=1}$   & $[\GeV^{-1}]$   & $2.07\pm 0.50 $     \\
    $w^{I=1}$   & $[\GeV^{-2}]$   & $-0.66\pm 0.10$     \\
    \hline
    $b^{I=0}$   & $[\GeV^{-1}]$   & $-0.82\pm 0.50$     \\
    $d^{I=0}$   & $[\GeV^{-1}]$   & $-1.95\pm 0.60$     \\
    $g^{I=0}$   & $[\GeV^{-1}]$   & $1.03\pm 0.90 $     \\
    $h^{I=0}$   & $[\GeV^{-1}]$   & $3.91\pm 0.90$     \\
    $w^{I=0}$   & $[\GeV^{-2}]$   & $-0.11\pm 0.40$     \\
    \hline
    $v_-$       & $[\GeV^{-1}]$   & $6.89\pm 0.19 $     \\
    $v_+$       & $[\GeV^{-1}]$   & $-1.98\pm 0.90$     \\
    \hline\hline
    $\chi^2_\mathrm{dof}$ & & 3.00                                                     \\
    \hline
  \end{tabular}
  % \begin{tabular}{c||D{.}{.}{12}|}
  %   \hline
  %   Fitting               & \multicolumn{1}{|c}{FIT~3$'$ (Bowen~1970 with $P_{01}$)} \\\hline  \hline
  %   $b^{I=1}$             & -0.39\pm 0.12                                            \\
  %   $d^{I=1}$             & 0.69\pm 0.18                                             \\
  %   $g^{I=1}$             & -1.07\pm 0.21                                            \\
  %   $h^{I=1}$             & 2.10\pm 0.50                                             \\
  %   $w^{I=1}$             & -0.66\pm 0.10                                            \\\hline
  %   $b^{I=0}$             & -0.80\pm 0.50                                            \\
  %   $d^{I=0}$             & -2.00\pm 0.60                                            \\
  %   $g^{I=0}$             & 1.00\pm 0.90                                             \\
  %   $h^{I=0}$             & -3.90\pm 0.90                                            \\
  %   $w^{I=0}$             & -0.10\pm 0.40                                            \\\hline
  %   $v_-$                 & 6.90\pm 0.19                                             \\
  %   $v_+$                 & -2.00\pm 0.90                                            \\
  %   \hline\hline
  %   $\chi^2_\mathrm{dof}$ & 3.00                                                     \\
  %   \hline
  % \end{tabular}
  \label{tab:FIT3_2}
\end{table}
\subsection{Quark condensate in $\SUN{3}$ symmetric baryonic matter}
% As seen in the previous section, we discussed the quark condensate including the strange quark component in symmetric nuclear matter consisting of the nucleons not including the strange quarks.
% In that sense, we discussed the SU(3) quark condensate in the SU(2) symmetric baryonic matter.
% It is also interesting to further extend the quark condensate in nuclear matter to that in the hypothetical hyperon matter and the SU(3) symmetric baryonic matter in order to discuss the SU(3) symmetry in the quark condensate.
% Note, however, that while the quark condensate in nuclear matter can be studied phenomenologically from the NG boson in heavy nuclei, as could be discussed from pionic atoms and pion-nuclei scattering, it may be difficult to make the phenomenological evaluation of the quark condensate in hyperon matter.
% In this section we derive the quark condensate in $\SUN{3}$ baryonic matter.
% The discussion here is in flavor $\SUN{3}$ limit.

In the previous section, we have discussed the quark condensate including the strange quark component in symmetric nuclear matter. 
{\red This is an SU(3) flavor extension of the quark condensate in nuclear matter. In the flavor symmetry point of view, it is also interesting to consider the SU(3) flavor extension of the matter.}
The nuclear matter consists of the nucleons without having explicit strange contents. In this sense, we have discussed an SU(3) quark condensate in the SU(2) symmetric baryonic matter. It may be also interesting to extend the discussion on the quark condensates in nuclear matter further to those in hypothetical hyperonic matter in order to discuss them in the aspect of the flavor SU(3) symmetry. 
{\red This kind of analyses might be interesting if one considers the explicit SU(3) breaking on the quark masses and the hadronic quantities such as the decay constants and masses. If one traces the SU(3) breaking effects on the in-medium quark condensate, the nuclear matter itself also can be a source of the SU(3) breaking.}
Note, however, that while the quark condensates in nuclear matter can be studied phenomenologically by the properties of the Nambu-Goldstone bosons in atomic nuclei as having done in pionic atoms and pion-nucleus scattering, the quark condensate in hyperonic matter would be rather academic due to the absence of hyperon matter in laboratories. 

Just as symmetric nuclear matter consists of the same number of protons and neutrons, we define SU(3) symmetric baryonic matter so as to consist of the same number of octet baryons with $J^p = {1/2}^+$, $p, n, \Lambda, \Sigma^+, \Sigma^0, \Sigma^-, \Xi^0$ and $\Xi^-$. We further consider $\Lambda$-hyperonic matter that contains only the $\Lambda$ hyperon, $\Sigma$-hyperonic matter which have the same number of $\Sigma^+$, $\Sigma^0$ and $\Sigma^-$, and $\Xi$-hyperonic matter which consists of the same numbers of $\Xi^0$ and $\Xi^-$. 
% We also consider symmetric baryonic matter consisting of the equal numbers of the octet baryons. 

The light quark condensate $\langle \bar uu +\bar dd \rangle$ in nuclear and hyperonic matter can be calculated in the same way as \Cref{sec:condensate} and are expressed in the linear density approximation by the isospin-averaged scattering amplitude of pion and the corresponding baryon in the soft limit like \Cref{eq:condensate}. The pion scattering amplitudes are calculated by chiral perturbation theory and expressed by the LECs. The relevant scattering amplitudes to the current calculation are shown in Appendix \ref{sec:appendix}. Taking the soft limit of the scattering amplitude, we obtain the quark condensate $\langle \bar uu +\bar dd \rangle$ in nuclear and hyperonic matter as
\begin{subequations}\label{eq:SU3cond}
  \begin{align}
    % \frac{\langle \bar uu +\bar dd \rangle^*_N}{\langle \bar uu + \bar dd \rangle_0}       & = 1 + \frac{4b_0+2b^{I=1}-2b^{I=0}}{F_\pi^2}\rho_B, \\
    % \frac{\langle \bar uu +\bar dd \rangle^*_\Lambda}{\langle \bar uu + \bar dd \rangle_0} & = 1 + \frac{\frac 83 b_0+ \frac 43 b^{I=1}}{F_\pi^2}\rho_B,    \\
    % \frac{\langle \bar uu +\bar dd\rangle^*_\Sigma}{\langle \bar uu + \bar dd \rangle_0}   & = 1 + \frac{4b^{I=1}}{F_\pi^2}\rho_B,      \\
    % \frac{\langle \bar uu +\bar dd \rangle^*_\Xi}{\langle \bar uu + \bar dd \rangle_0}     & = 1 + \frac{2b^{I=1}+2b^{I=0}}{F_\pi^2}\rho_B,
   \frac{\langle \bar uu +\bar dd \rangle^*_N}{\langle \bar uu + \bar dd \rangle_0}       & = 1 + \frac{4b_0+2b_D+2b_F}{F_\pi^2}\rho_B, \\
   \frac{\langle \bar uu +\bar dd \rangle^*_\Lambda}{\langle \bar uu + \bar dd \rangle_0} & = 1 + \frac{4b_0+\frac43 b_D}{F_\pi^2}\rho_B,    \\
   \frac{\langle \bar uu +\bar dd\rangle^*_\Sigma}{\langle \bar uu + \bar dd \rangle_0}   & = 1 + \frac{4b_0+4b_D}{F_\pi^2}\rho_B,      \\
   \frac{\langle \bar uu +\bar dd \rangle^*_\Xi}{\langle \bar uu + \bar dd \rangle_0}     & = 1 + \frac{4b_0+2b_D-2b_F}{F_\pi^2}\rho_B,
  \end{align}
\end{subequations}
where $\rho_B$ is the density of the baryon number in each baryonic matter, and we write these expressions in terms of the original LECs appearing in the Lagrangian in order to make the SU(3) flavor structure clear. The relation to the LECs $b^{I=1}$ and $b^{I=0}$ are given in \Cref{eq:bparas}. Similarly, the quark condensate $\langle \bar uu +\bar ss \rangle$ in hyperonic matter is obtained by the soft limit of the isospin averaged kaon-hyperon scattering amplitude in the linear density approximation and expressed by the LECs as
\begin{subequations}
  % \begin{align}
  %   \frac{\langle \bar uu +\bar ss \rangle^*_N}{\langle \bar uu + \bar ss \rangle_0}       & = 1 + \frac{3b^{I=1}+b^{I=0}}{F_K^2}\rho_B, \\
  %   \frac{\langle \bar uu +\bar ss \rangle^*_\Lambda}{\langle \bar uu + \bar ss \rangle_0} & = 1 + \frac{\frac 23 b_0+ \frac{10}3 b^{I=1}}{F_K^2}\rho_B,   \\
  %   \frac{\langle \bar uu +\bar ss \rangle^*_\Sigma}{\langle \bar uu + \bar ss \rangle_0}  & = 1 + \frac{2b_0+2b^{I=1}}{F_K^2}\rho_B,     \\
  %   \frac{\langle \bar uu +\bar ss \rangle^*_\Xi}{\langle \bar uu + \bar ss \rangle_0}     & = 1 + \frac{2b_0+3b^{I=1}-b^{I=0}}{F_K^2}\rho_B,
  % \end{align}
 \begin{align}
   \frac{\langle \bar uu +\bar ss \rangle^*_N}{\langle \bar uu + \bar ss \rangle_0}       & = 1 + \frac{4b_0+3b_D-b_F}{F_K^2}\rho_B, \\
   \frac{\langle \bar uu +\bar ss \rangle^*_\Lambda}{\langle \bar uu + \bar ss \rangle_0} & = 1 + \frac{4b_0+\frac{10}3 b_D}{F_K^2}\rho_B,   \\
   \frac{\langle \bar uu +\bar ss \rangle^*_\Sigma}{\langle \bar uu + \bar ss \rangle_0}  & = 1 + \frac{4b_0+2b_D}{F_K^2}\rho_B,     \\
   \frac{\langle \bar uu +\bar ss \rangle^*_\Xi}{\langle \bar uu + \bar ss \rangle_0}     & = 1 + \frac{4b_0+3b_D+b_F}{F_K^2}\rho_B.
 \end{align}
\end{subequations}
The quark condensates in the $\SUN{3}$ symmetric baryonic matter are also obtained as:
\begin{align}\label{eq:SU3cond2}
  \frac{\langle \bar uu +\bar dd \rangle^*_B}{\langle \bar uu + \bar dd \rangle_0} & = \frac{1}{8}\qty[2\frac{\langle \bar uu +\bar dd \rangle^*_N}{\langle \bar uu + \bar dd \rangle_0} + \frac{\langle \bar uu +\bar dd \rangle^*_\Lambda}{\langle \bar uu + \bar dd \rangle_0} + 3\frac{\langle \bar uu +\bar dd \rangle^*_\Sigma}{\langle \bar uu + \bar dd \rangle_0} + 2\frac{\langle \bar uu +\bar dd \rangle^*_\Xi}{\langle \bar uu + \bar dd \rangle_0}] \nonumber  \\
  &   %= 1 + \frac43 \frac{b_0 + 2 b^{I=1}}{F_\pi^2} \rho_B 
  = 1 + \frac{4b_0 + \frac83 b_D}{F_\pi^2} \rho_B, \\
  \frac{\langle \bar uu +\bar ss \rangle^*_B}{\langle \bar uu + \bar ss \rangle_0} & = \frac{1}{8}\qty[2\frac{\langle \bar uu +\bar ss \rangle^*_N}{\langle \bar uu + \bar ss \rangle_0} + \frac{\langle \bar uu +\bar ss \rangle^*_\Lambda}{\langle \bar uu + \bar ss \rangle_0} + 3\frac{\langle \bar uu +\bar ss \rangle^*_\Sigma}{\langle \bar uu + \bar ss \rangle_0} + 2\frac{\langle \bar uu +\bar ss \rangle^*_\Xi}{\langle \bar uu + \bar ss \rangle_0}] \nonumber \\
  &= 1 + \frac{4b_0 + \frac83 b_D}{F_K^2} \rho_B .
\end{align}
%where we relabel the quark condensate in nuclear matter $\langle \rangle^*$ as $\langle \rangle^*_N$ to distinguish it from that in other baryonic matter.
The SU(3) quark condensate in the $\SUN{3}$ symmetric baryonic matter is calculated as
\begin{align} \label{eq:SU3cond3}
  \frac{\langle \bar uu +\bar dd +\bar ss\rangle^*_B}{\langle \bar uu + \bar dd +\bar ss\rangle_0} = \frac{2}{3} \qty(\frac{\langle \bar uu +\bar ss \rangle^*_B}{\langle \bar uu + \bar ss \rangle_0} + \frac{1}{2}\frac{\langle \bar uu +\bar dd \rangle^*_B}{\langle \bar uu + \bar dd \rangle_0})
  &= 1 + \frac{4b_0 + \frac83 b_D}{F^2} \rho_B,
\end{align}
where we assume the flavor symmetry for the in-vacuum condensates $\langle \bar uu\rangle_0 = \langle \bar dd\rangle_0 = \langle \bar ss\rangle_0$ and the meson decay constants $F=F_\pi = F_K$. As one expects, the slope parameters of Eqs.~\eqref{eq:SU3cond}, \eqref{eq:SU3cond2} and \eqref{eq:SU3cond3} should be equivalent according to the flavor symmetry because the matter is flavor-symmetric.  
% which seems to be valid according to the lattice QCD calculation \cite{McNeile:2012xh}.

As we have seen in the previous section, to evaluate the quark condensate $\langle \bar uu + \bar ss \rangle$ in the nuclear matter, we just need the two-parameters $b^{I=1}$ and $b^{I=0}$, which can be fixed by the $K^+N$ scattering.
On the other hand, for other cases we need to know the value of $b_0$. Here we determine it by \Cref{eq:assumption_b0} with $c_1=-0.78$~GeV$^{-1}$. In the following we use the LECs $b^{I=1}$ and $b^{I=0}$ determined in FIT~2 as an example. 
% Then we can obtain the LECs $b_0$, $b_D$ and $b_F$ shown in \Cref{tab:SU3LECs2}.
% \begin{table}[H]
%     \centering
%     \caption{$\SUN{3}$ low-energy constants obtained by the fitting of $KN$ scattering with the assumption (\ref{eq:assumption_b0}) and other calculations are shown in unit of $\GeV^{-1}$.}
%     \begin{tabular}{l|rrrrr}
%         \hline
%         LEC   & FIT~1   & FIT~2   & FIT~3   & FIT~4   & FIT~3'  \\
%         \hline\hline
%         $b_0$ & $-1.99$ & $0.58$  & $0.54$  & $0.99$  & $-0.91$ \\
%         $b_D$ & $0.92$  & $-1.69$ & $-0.65$ & $-2.08$ & $0.53$  \\
%         $b_F$ & $1.67$  & $-0.87$ & $-1.82$ & $-1.30$ & $-0.09$ \\
%         \hline
%     \end{tabular}
%     \label{tab:SU3LECs2}
% \end{table}

\begin{figure}[H]
  \centering
  \includegraphics[width=10cm,clip]{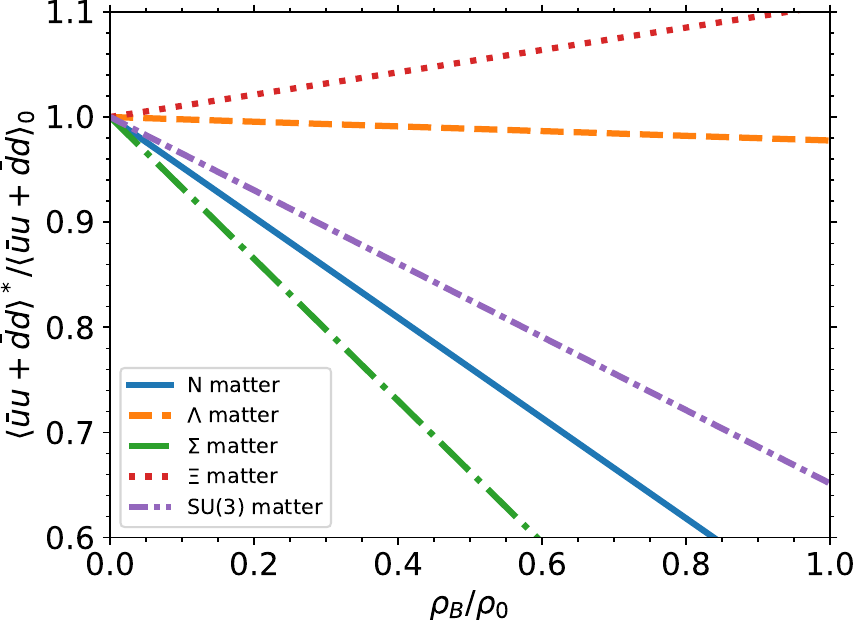}
  \caption{Baryon density dependence of $\langle \bar uu +\bar dd \rangle$ in nuclear matter, hyperonic matter and the $\SUN{3}$ symmetric baryonic matter. The LECs of FIT~2 are used and $b_0$ is fixed by \Cref{eq:assumption_b0} with $c_1 = - 0.78\ \GeV^{-1}$. We use $F_\pi=93$ MeV and $\rho_0 = 0.17$ fm$^{-3}$.}
  \label{fig:uu_plus_dd_cond_B}
\end{figure}

Firstly, we plot the behavior of $\langle \bar uu +\bar dd \rangle$ in nuclear matter, hyperon matter and the $\SUN{3}$ symmetric baryonic matter in \Cref{fig:uu_plus_dd_cond_B}.
This figure shows that the SU(3) flavor symmetry breaking for the baryonic matter since the condensate $\langle \bar uu +\bar dd \rangle$ has no strange components but the hyperonic matter contains the strange quarks.
The relative amount of the up and down quarks in the hyperonic matter is less than in nuclear matter, so the condensate in the hyperonic matter is expected to decrease less than that in nuclear matter.
Figure \ref{fig:uu_plus_dd_cond_B} shows that the quark condensates in $\Lambda$-matter and $\Xi$-matter increase in magnitude, while the quark condensate in $\Sigma$ hyperonic matter decreases more than that in nuclear matter. 
On the other hand, the condensate in the $\SUN{3}$ symmetric baryonic matter is reduced but not more than that in nuclear matter, this is an expected behavior.

\begin{figure}[H]
  \centering
  \includegraphics[width=10cm,clip]{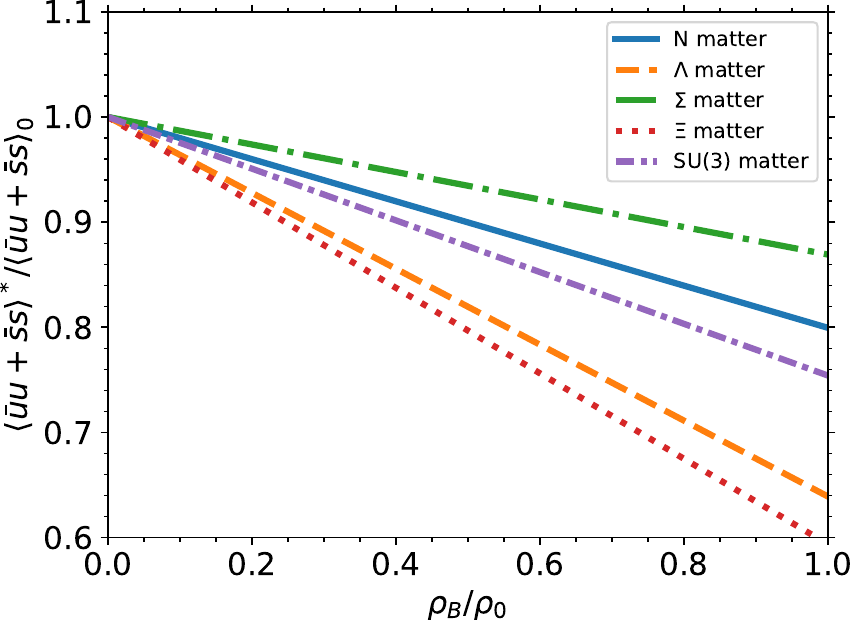}
  \caption{Same as \Cref{fig:uu_plus_dd_cond_B}, but for $\langle \bar uu +\bar ss \rangle$.}
  \label{fig:uu_plus_ss_cond_B}
\end{figure}

Next, we plot the density dependence of $\langle \bar uu +\bar ss \rangle$ in nuclear matter, hyperonic matter and the $\SUN{3}$ symmetric baryonic matter shown in \Cref{fig:uu_plus_ss_cond_B}.
% \Cref{fig:uu_plus_ss_cond_B} show that the behavior of the quark condensate in each hyperon matter and the $\SUN{3}$ symmetric baryonic matter.
The calculation shows that the quark condensates $\langle \bar uu +\bar ss \rangle$ in $\Lambda$ hyperonic matter and $\Xi$ hyperonic matter are reduced compared to the quark condensate in nuclear matter, but the condensate in $\Sigma$-matter is reduced less than that in nuclear matter.
Thus, since hyperonic matter contains strange quarks, one expects that the quark condensate with strange components in hyperonic matter would be reduced compared to quark condensate in nuclear matter, but this is not necessarily the case.
On the other hand, the condensate in the $\SUN{3}$ symmetric baryonic matter is reduced compared to that in nuclear matter, this is also an expected behavior.

\subsection{Wave function renormalization of in-medium kaon}\label{sec:WFR}
The wave function renormalization of the NG bosons in the nuclear medium has been investigated as one of the important in-medium modifications of the hadron properties, for instance, in Refs.~\cite{Jido:2000bw,Kolomeitsev:2002gc,Jido:2007yt,Jido:2008bk,Goda:2013npa,Jido:2016jnl,Aoki:2017hel}. References~\cite{Jido:2000bw,Jido:2007yt} pointed out that the pion wave function renormalization in the nuclear medium is responsible for the in-medium change of the pion decay constant. In Ref.~\cite{Kolomeitsev:2002gc}, the wave function renormalization for the in-medium pion was discussed to explain the missing repulsion of the in-medium $\pi N$ scattering length. Reference~\cite{Aoki:2017hel} calculated the wave function renormalization for the in-medium kaon using the $K^+ N$ amplitude described by chiral dynamics and found that the leading order analysis with the Weinberg-Tomozawa interaction suggested 8 \% enhancement of the wave function normalization factor at the normal nuclear density and full calculations provided about 2 to 6\% enhancement depending on the kaon momentum. This indicates that the $K^+ N$ interaction may get enhanced about several percent in nuclear matter. This is partially consistent with the phenomenological finding of the enhancement of the $K^+N$ elastic scattering amplitude in nucleus~\cite{Bugg:1968zz,Siegel:1985hp,Weise:1989um,Weiss:1994kt,Friedman:1997as}.

Here we update the study of Ref.~\cite{Aoki:2017hel} by using the $K^+ N$ scattering amplitudes constructed using more general terms in chiral perturbation theory and determined by wider fitting procedures. According to Ref.~\cite{Aoki:2017hel}, the wave function renormalization factor $Z_K$ for the in-medium kaon is obtained by using the optical potential for a kaon in nuclear matter $V_\mathrm{opt}$ as
\begin{align}
  Z_K & \equiv 1 + \frac{M_K}{\omega_K}\eval{\pdv{V_\mathrm{opt}}{\omega_K^*}}_{\omega_K^* = \omega_K},
\end{align}
where $\omega_K$ is the kaon energy. In the linear density approximation the optical potential is given by the $KN$ scattering amplitude as 
\begin{align}
  2M_K V_\mathrm{opt}(\omega_K) = \frac{\rho}{2M_N} T_{K N}(\omega_K).
\end{align}
We calculate the wave function renormalization for the in-medium kaon using the $K^+ N$ scattering amplitudes constructed in the previous section. 

% \begin{align}
%  V_\mathrm{opt}(\omega_K^*) & = V_\mathrm{opt}(\omega_K) + (\omega_K^*-\omega_K)\eval{\pdv{V_\mathrm{opt}}{\omega_K^*}}_{\omega_K^* = \omega_K}  + \dots\nonumber                     \\
%                             & \simeq V_\mathrm{opt}(\omega_K) + \frac{M_K}{\omega_K} V_\mathrm{opt}(\omega_K)\eval{\pdv{V_\mathrm{opt}}{\omega_K^*}}_{\omega_K^* = \omega_K}\nonumber \\
%  %  & = \qty(1 + \frac{M_K}{\omega_K}\eval{\pdv{V_\mathrm{opt}}{\omega_K^*}}_{\omega_K^* = \omega_K})V_\mathrm{opt}(\omega_K)
%                             & \equiv Z_K V_\mathrm{opt}(\omega_K),
%\end{align}

\begin{figure}[H]
  \centering
  \includegraphics[width=0.6\textwidth]{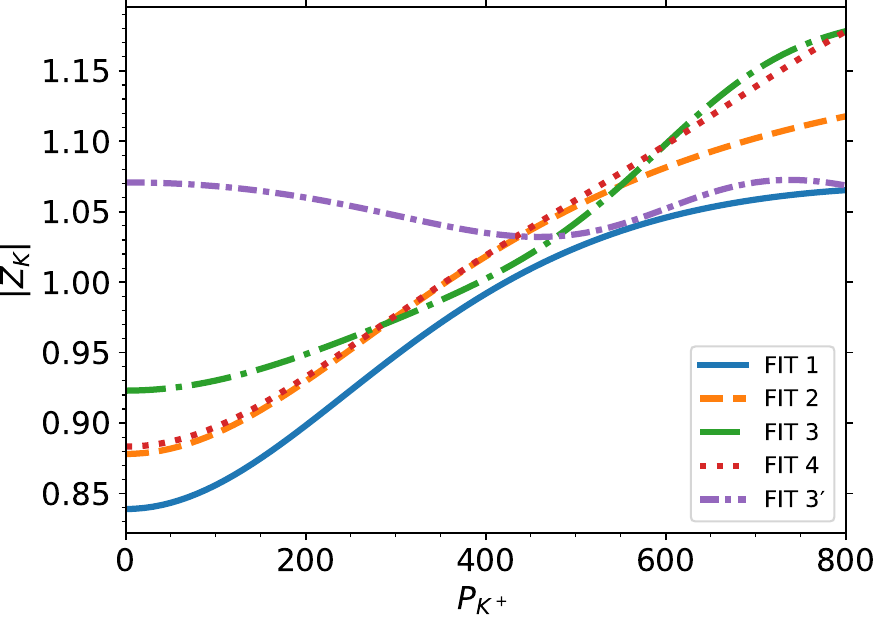}
  \caption{Momentum dependence of the absolute value of the wave function renormalization factor $Z_K$ for kaon at the normal nuclear density.}
  \label{fig:wavefunction_renormalization}
\end{figure}

\begin{figure}[H]
  \centering
  \includegraphics[width=0.6\textwidth]{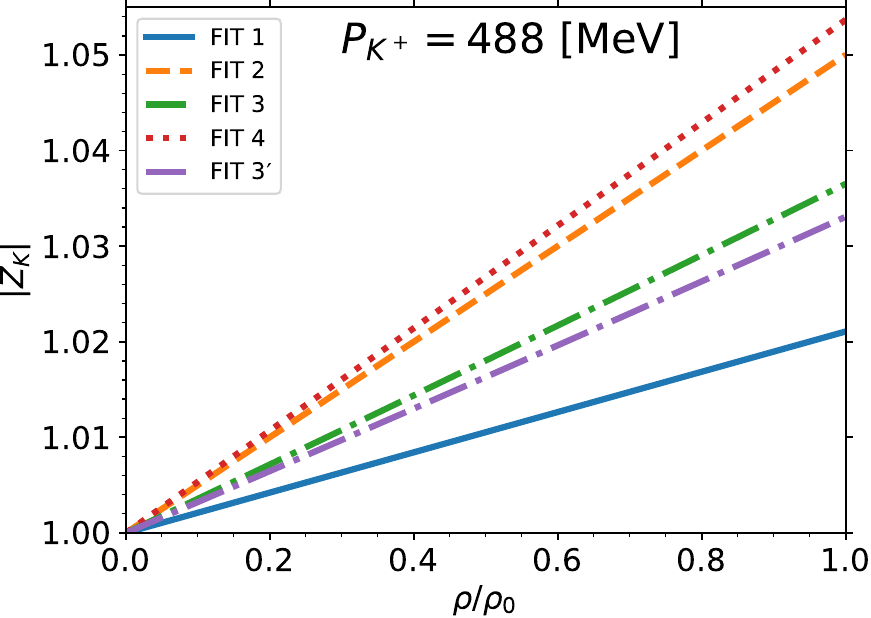}
  \caption{Density dependence of the absolute value of the wave function renormalization factor $Z_K$ at the $P_{K^+} = 488\MeV/c$.}
  \label{fig:wavefunction_renormalization2}
\end{figure}

The wave function renormalization factor $Z_K$ at the normal nuclear density is shown in \Cref{fig:wavefunction_renormalization} as a function of the momentum of kaon in nuclear matter $P_{K^+}$. We find in the figure that the momentum dependence of wave function renormalization factors obtained by FITs 1 to 4 is qualitatively consistent with each other and monotonically increases with respect to $P_{K^+}$, while $Z_K$ with FIT~3$'$ is almost independent of $P_{K^+}$ and gives almost 6\% enhancement. We show the linear density dependence of the wave function renormalization factor $Z_K$ at $P_{K^+} = 488\MeV/c$ in \Cref{fig:wavefunction_renormalization2}. The enhancement of the wave function renormalization factors is found to be around 2\% to 5\% depending on the fitting procedures. This result is consistent with the previous study. In the case of the in-medium pion, the wave function renormalization factor is enhanced by $40\%$ at the normal nuclear density \cite{Goda:2013npa}.
Compared to the case of the in-medium pion, our calculation gives a smaller enhancement at the normal nuclear density.

\section{Summary}\label{sec:conclusion}
We have investigated the $K^+N$ scattering amplitude using chiral perturbation theory in order to estimate the in-medium quark condensate with strange quarks. The in-medium quark condensate is calculated based on the correlation function approach. There the in-medium quark condensate with the strange quarks is given by the correlation function of the pseudoscalar fields with kaon quantum number in nuclear matter at the soft limit. In the linear density approximation, the in-medium correlation function is reduced to the product of the $KN$ scattering amplitude and the nuclear density. We utilize chiral perturbation theory to describe the $KN$ scattering amplitude. It is good that the amplitude in chiral perturbation theory is described by an analytic function and can be analytically continued to the soft limit. 

We have determined the low energy constants (LECs) of the SU(3) chiral perturbation theory appearing in the $KN$ scattering amplitude by the existing scattering data. The scattering amplitudes has been calculated up to the next-to-leading order in chiral perturbation theory and in addition we have also included the strange quark mass dependent terms of the next-to-next-to-leading order in order to improve extrapolation to the strange sector. The LECs appearing here characterize the interaction between $K^+$ and $N$. We have performed several fitting procedures for the LECs using the experimental data of the $K^+ p$ differential cross section, the $K^+ n\to K^0 p$ charge exchange differential cross sections, and $I=1$ and $I=0$ total cross sections. For the experimental data of the $I=0$ total cross section, we take two choices because two data sets look inconsistent. In addition, we have further choices to include a broad resonance state with $I=0$ around $P_\mathrm{lab} = 600\MeV/c$, which was proposed in Ref.~\cite{Aoki:2017hel}, or not. We have obtained such a nice amplitude for $I=1$ that it reproduces the experimental data below $P_\mathrm{lab} = 800\MeV$ almost perfectly. For the $I=0$ amplitude, we have used the $I=0$ total cross section and the differential cross section of the $K^+n\to K^0 p$ to determine the LECs in the $I=0$ amplitude. We have found that the scattering data for $I=0$ are also reproduced well but the LECs are not uniquely determined and depend on the fitting procedures. In addition it has turned out that the differential cross section of the $K^+n$ elastic scattering are not reproduced even though the isospin symmetry should fix the $K^+n$ amplitude from the $K^+ p \to K^+ p$ and $K^+ n \to K^0 p$ amplitudes.

With the determined LECs, we have discussed the behavior of the in-medium quark condensate with strange components in the linear density approximation. We have found that the slope parameter of the linear density is dependent on the fitting procedures. This implies that the current $K^+N$ experiment data especially in low energies do not have enough accuracy to fix the LECs. Some parameter sets provide consistent results of the slope parameter with other determinations of the LECs such as those based on the baryon masses in lattice calculations for various quark masses. We have also calculated the quark condensates, $\langle \bar uu +\bar dd \rangle$ and $\langle \bar uu +\bar ss \rangle$, in hyperonic matter and the $\SUN{3}$ symmetric baryonic matter in the aspect of the flavor symmetry. Moreover, we have calculated $\langle \bar uu +\bar dd +\bar ss\rangle$ in the $\SUN{3}$ symmetric baryonic matter and obtained the $25\%$ restoration of the chiral symmetry in the case of SU(3) with our fitted LECs. This result is consistent with the case of the SU(2) condensate in nuclear matter. We have calculated the wave function renormalization factor using the obtained $T$-matrix of $KN$. 
%The dependence of the momentum of kaon in nuclear matter $P_{K^+}$ of the wave function renormalization factor $Z_K$ at the normal nuclear density depends on the fitting procedures. 
In any FITs, the wave function renormalization factor for in-medium kaon with an intermediate momentum such as $P_{K^+} = 488\MeV/c$ increases as the density increases, but the enhancement is not as large as that for in-medium pion.

In conclusion, 
{\red thanks to good accuracy and wide range of the $K^+p$ elastic scattering data, the $KN$ scattering amplitude with $I=1$ is well controlled in chiral perturbation theory. With this success, the correlation function approach with the linear density approximation has worked well to see qualitative feature of the in-medium strange quark condensate. Nevertheless,}
we emphasize that in order to determine the behavior of the in-medium quark condensate with strange quarks more accurately, it is important to determine the $K^+N$ scattering amplitudes in the energies much lower than $P_\mathrm{lab} = 400\MeV/c$ where the amplitude may be free from the effect of the possible resonance state.

\section*{Acknowledgements}
We would liket to thank Dr.~K.~Aoki for his giving us the resonance amplitudes. 
The work of Y.I.~was partly supported by Grants-in-Aid for Scientific Research from JSPS (20J20598).
The work of D.J.~was partly supported by Grants-in-Aid for Scientific Research from JSPS (JP21K03530 and JP22H04917).

\appendix
\section{Meson-baryon scattering $T$-matrices for the quark condensates} \label{sec:appendix}
In this section, we give the list of the meson-baryon scattering $T$-matrices relevant to the calculation of the in-medium quark condensates. As seen in \Cref{eq:condensate}, we need the $T$-matrices of the meson-baryon scattering in the soft-limit.
As discussed in Section \ref{sec:ChPT}, the relevant terms in the $T$-matrix to the quark condensate are the terms involving the LECs $b_0,b_D$ and $b_F$ which appear in the next-to-the leading order of chiral Lagrangian \Cref{eq:NLO}.
%Therefore, we list the $T$-matrices containing the LECs $b_0,b_D$ and $b_F$.

For the evaluation of the in-medium condensate $\langle \bar uu +\bar dd \rangle$, we use the $T$-matrices of the pion-baryon:
\begin{subequations}
  \begin{align}
    T_{\pi^0 p}        & = \frac{2B_0 m (4b_0+2b_D+2b_F)}{F_\pi^2} \times 2M_N,       \\
    T_{\pi^0 n}        & = \frac{2B_0 m (4b_0+2b_D+2b_F)}{F_\pi^2} \times 2M_N,       \\
    T_{\pi^0 \Lambda}  & = \frac{2B_0 m (4b_0+\frac43 b_D)}{F_\pi^2}    \times 2M_\Lambda, \\
    T_{\pi^0 \Sigma^+} & = \frac{2B_0 m (4b_0+4b_D)}{F_\pi^2}      \times 2M_\Sigma,  \\
    T_{\pi^0 \Sigma^-} & = \frac{2B_0 m (4b_0+4b_D)}{F_\pi^2}      \times 2M_\Sigma,  \\
    T_{\pi^0 \Sigma^0} & = \frac{2B_0 m (4b_0+4b_D)}{F_\pi^2}      \times 2M_\Sigma,  \\
    T_{\pi^0 \Xi^0}    & = \frac{2B_0 m (4b_0+2b_D-2b_F)}{F_\pi^2} \times 2M_\Xi,     \\
    T_{\pi^0 \Xi^-}    & = \frac{2B_0 m (4b_0+2b_D-2b_F)}{F_\pi^2} \times 2M_\Xi.
  \end{align}
  % \begin{align}
  %   T_{\pi^0 p}        & = \frac{4B_0 m (2b_0+b_D+b_F)}{F_\pi^2} \times 2M_N,       \\
  %   T_{\pi^0 n}        & = \frac{4B_0 m (2b_0+b_D+b_F)}{F_\pi^2} \times 2M_N,       \\
  %   T_{\pi^0 \Lambda}  & = \frac{8B_0 m (3b_0+b_D)}{3F_\pi^2}    \times 2M_\Lambda, \\
  %   T_{\pi^0 \Sigma^+} & = \frac{8B_0 m (b_0+b_D)}{F_\pi^2}      \times 2M_\Sigma,  \\
  %   T_{\pi^0 \Sigma^-} & = \frac{8B_0 m (b_0+b_D)}{F_\pi^2}      \times 2M_\Sigma,  \\
  %   T_{\pi^0 \Sigma^0} & = \frac{8B_0 m (b_0+b_D)}{F_\pi^2}      \times 2M_\Sigma,  \\
  %   T_{\pi^0 \Xi^0}    & = \frac{4B_0 m (2b_0+b_D-b_F)}{F_\pi^2} \times 2M_\Xi,     \\
  %   T_{\pi^0 \Xi^-}    & = \frac{4B_0 m (2b_0+b_D-b_F)}{F_\pi^2} \times 2M_\Xi.
  % \end{align}
\end{subequations}

For $\langle \bar uu +\bar ss \rangle$, we use the $T$-matrices of the kaon-baryon:
\begin{subequations}
  \begin{align}
    T_{K^+p}        & = \frac{B_0(m+m_s)(4b_0+4b_D)}{F_K^2}       \times 2M_N,       \\
    T_{K^+n}        & = \frac{B_0(m+m_s)(4b_0+2b_D-2b_F)}{F_K^2}  \times 2M_N,       \\
    T_{K^+\Lambda}  & = \frac{B_0(m+m_s)(4b_0+\frac {10}3 b_D)}{F_K^2}    \times 2M_\Lambda, \\
    T_{K^+\Sigma^+} & = \frac{B_0(m+m_s)(4b_0+2b_D+2b_F)}{F_K^2}  \times 2M_\Sigma,  \\
    T_{K^+\Sigma^-} & = \frac{B_0(m+m_s)(4b_0+2b_D-2b_F)}{F_K^2}  \times 2M_\Sigma,  \\
    T_{K^+\Sigma^0} & = \frac{B_0(m+m_s)(4b_0+2b_D)}{F_K^2}      \times 2M_\Sigma,  \\
    T_{K^+\Xi^0}    & = \frac{B_0(m+m_s)(4b_0+2b_D+2b_F)}{F_K^2}  \times 2M_\Xi,     \\
    T_{K^+\Xi^-}    & = \frac{B_0(m+m_s)(4b_0+4b_D)}{F_K^2}       \times 2M_\Xi.
    % [\im\mathcal L_{\bar pp K^0\bar K^0}]               & = -\frac{2\im B_0(m+m_s)(2b_0+b_D-b_F)}{F_K^2} \\
    % [\im\mathcal L_{\bar nn K^0\bar K^0}]               & = -\frac{4\im B_0(m+m_s)(b_0+b_D)}{F_K^2}      \\
    % [\im\mathcal L_{\bar \Lambda\Lambda K^0\bar K^0}]   & = -\frac{2\im B_0(m+m_s)(6b_0+5b_D)}{3F_K^2}   \\
    % [\im\mathcal L_{\bar \Sigma^+\Sigma^+ K^0\bar K^0}] & = -\frac{2\im B_0(m+m_s)(2b_0+b_D-b_F)}{F_K^2} \\
    % [\im\mathcal L_{\bar \Sigma^0\Sigma^0 K^0\bar K^0}] & = -\frac{4\im B_0(m+m_s)(b_0+b_D)}{F_K^2}      \\
    % [\im\mathcal L_{\bar \Sigma^-\Sigma^- K^0\bar K^0}] & = -\frac{2\im B_0(m+m_s)(2b_0+b_D+b_F)}{F_K^2} \\
    % [\im\mathcal L_{\bar \Xi^0\Xi^0 K^0\bar K^0}]       & = -\frac{4\im B_0(m+m_s)(b_0+b_D)}{F_K^2}      \\
    % [\im\mathcal L_{\bar \Xi^-\Xi^- K^0\bar K^0}]       & = -\frac{2\im B_0(m+m_s)(2b_0+b_D+b_F)}{F_K^2}
  \end{align}
\end{subequations}

\bibliographystyle{apsrev4-2}
\bibliography{2308-018-3D-DaisukeJido.bib}

\end{document}